\documentclass[a4paper,11pt]{article}	
\pdfoutput=1 
\usepackage{jheppub} 
\usepackage{mathstyle} 
\usepackage[final]{showkeys}
\usepackage{makeidx}
\usepackage{mathrsfs}

\def\IC{\mathbb{C}}

\def\IH{\mathbb{H}}

\def\IN{\mathbb{N}} 
\def\IP{\mathbb{P}}

\def\IZ{\mathbb{Z}}

\renewcommand{\ri}{{\rm i}}

\newcommand{\CA}{\mc{A}}

\newcommand{\CD}{\mc{D}}

\newcommand{\CF}{\mc{F}} 
 
\newcommand{\CH}{\mc{H}} 
\newcommand{\CM}{\mc{M}}
\newcommand{\CN}{\mc{N}}
\newcommand{\CO}{\mc{O}}
\newcommand{\CS}{\mc{S}}

\newcommand{\SD}{\ms{D}}

\newcommand{\Disc}{\text{Disc}}
\newcommand{\Cont}{\text{Cont}}

\newcommand{\tsp}[1]{#1^T}

\newcommand{\at}{t}

\renewcommand{\jjj}[1]{}

\makeindex

\title{\boldmath Modularity of resurgent topological string}%

\author[a]{Gengbei Guo}%
\author[b]{Jiashen Chen}%
\author[a]{Jie Gu}%
\affiliation[a]{School of Physics and Shing-Tung Yau Center\\
  Southeast University, Nanjing 210096, China}%
\affiliation[b]{Interdisciplinary Center for Theoretical Study,
University of Science and Technology of China, Hefei, Anhui 230026, China}

\emailAdd{gengbei-guo@seu.edu.cn}
\emailAdd{ricepud@mail.ustc.edu.cn}
\emailAdd{jie-gu@seu.edu.cn}

\abstract{Topological string free energy has a rich collection of
  non-perturbative contributions which are labeled by D-brane charge
  vectors, and the associated Stokes constants are conjectured to
  coincide with BPS or DT invariants, i.e.~D-brane multiplicities.  In
  this paper, we provide additional evidence to this conjecture by
  studying modular properties of non-perturbative contributions.  We
  argue using resurgence theory that 
  non-perturbative contributions form orbits of local monodromy group
  induced by singular points inside a stability chamber, and that the
  associated Stokes constants must be the same across the orbits. In
  some examples, this allows generation of infinitely many Stokes
  constants, which reproduce the entire BPS spectrum.  In addition,
  following \cite{Douaud:2026qfo}, we also show that generators of
  Stokes transformations of non-holomorphic partition function satisfy
  Lie brackets of the Kontsevich-Soibelman Lie algebra, making it
  possible to identify the global Stokes transformation with the
  Kontsevich-Soibelman wall-crossing invariant.
  \\

\begin{center}
These authors contributed equally to this work.\end{center}}

\begin{document}
\maketitle
\flushbottom


\section{Introduction}

One of the central ideas in string theory is that perturbative string
theory needs to be completed by non-perturbative contributions, which
should come from higher dimensional objects such as D-branes 
\cite{Polchinski:1994fq}.  An ideal laboratory to check this idea is
topological string theory, a simplified version of string theory.  The
perturbative free energy of topological string
\begin{equation}
  \CF(t,g_s) = \sum_{g=0}^\infty \CF_g(t) g_s^{2g-2} 
\end{equation}
can be defined rigorously in mathematics in terms of the Gromov-Witten
theory, have a nice integral structure encoded in the Gopakumar-Vafa
invariants \cite{Gopakumar:1998ii,Gopakumar:1998jq}, and can be
calculated by efficient methods such as holomorphic anomaly equations
\cite{Bershadsky:1993cx,Bershadsky:1993ta}, so that the perturbative
series can be computed to very higher orders.  It can be verified that
the perturbative free energy of topological string is indeed a
divergent power series as the coefficients $\CF_g(t)$ grow factorially
fast with respect to $g$
\cite{Marino:2006hs,Marino:2007te,Marino:2008ya}, so that it requires
non-perturbative corrections of order $\CO(\re^{-1/g_s})$.
Furthermore, given rich perturbative data, it is natural to apply the
resurgence theory to explore possible non-perturbative corrections.

Resurgence theory is a powerful mathematical framework that studies
the relationship between perturbative asymptotic series and its
non-perturbative corrections \cite{Ecalle} \footnote{See
  \cite{Marino:2012zq,Mitschi:2016fxp,Aniceto:2018bis,Dunne:2025mye}
  as well as \cite[Appendix]{Douaud:2026qfo} for pedagogical
  lectures.}.  According to the resurgence theory, non-perturbative
corrections should arise in the form of transseries, a power series
weighted by an exponential suppression factor $\re^{-\CA/g_s}$
characterised by the action $\CA$, and non-perturbative contributions
are intimately related to the perturbative series by the means of
Stokes transformation.  As a consequence, vital non-perturbative
information can be extracted from the perturbative series itself by
the so-called resurgence relation.  It is then an interesting question
if non-trivial non-perturbative contributions can be derived for
topological string using the resurgence theory, and if these
non-perturbative contributions are related to D-branes in some way.

The program to study non-perturbative corrections to topological
string free energy was initiated in
\cite{Marino:2006hs,Marino:2007te,Marino:2008ya}, followed for
instance by \cite{Pasquetti2010,Aniceto2012}. \footnote{See
  \cite{Marino:2024tbx} for review and recent progress.}
For the simple examples of conifold free energies and resolved
conifold, the non-perturbative free energies are found to be of a
particular simple form \cite{Pasquetti2010} known as the
Pasquetti-Schiappa instanton amplitude.  In addition, by analysing the
free energies near the large radius point, it was discovered in
\cite{Pasquetti2010,Aniceto2012,Couso-Santamaria:2016vwq,Gu:2023mgf}
that for every non-vanishing GV invariant $n_{0,\md{d}}$ of genus zero
supported on a 2-cycle $\md{d}$ in the Calabi-Yau threefold, there is
an family of non-perturbative sectors whose actions are given by
\begin{equation}
  \CA_{(\md{d},m)} = 2\pi\md{d}\cdot\md{t} + 4\pi^2 \ri m,\quad m\in \IZ
\end{equation}
where $\md{t}$ is the local flat coordinate near the LR point on the
moduli space, and whose non-perturbative contributions are of the
Pasquetti-Schiappa form.  Furthermore, the coefficients of the Stokes
transformations of these non-perturbative contributions from the
perturbative free energy, the Stokes constants, are nothing else but
the GV invariant $n_{0,\md{d}}$.  As the GV invariant $n_{0,\md{d}}$
counts the multiplicity of the BPS states of $m$ D0-branes bound to
D2-branes wrapping the curve class $\md{d}$ in $X$ in type IIA
superstring, this is the first indication that the nonperturbative
contributions of topological string originate possibly from D-branes.

The appearance of GV invariants may not seem surprising as they are
already encoded in the perturbative free energies themselves. However,
topological string has indeed richer non-perturbative contributions
beyond just GV invariants. Already in \cite{Pasquetti2010} the studies
of the conifold free energies indicate that there are non-perturbative
contributions from D4- or D6-branes which become massless at the
conifold singular points of the moduli space.  A crucial development
in this direction was the closed form formulas of non-perturbative
contributions in any possible non-perturbative sector
\cite{Gu:2022sqc,Gu:2023mgf}, based on the idea
\cite{Couso-Santamaria2015,Couso-Santamaria2016} that non-perturbative
contributions are transseries solutions to the holomorphic anomaly
equations. \footnote{In the case of local Calabi-Yau threefolds, the
  same formulae were also derived as symplectic transformation of the
  simple non-perturbative contributions in the Pasquetti-Schiappa form
  \cite{Marino:2024yme,Marino:2024tbx}.}
Along the way, arguments and evidence were provided
\cite{Drukker:2011zy,Couso-Santamaria2015,Couso-Santamaria2016} that
the action for any non-perturbative sector was a linear combination of
the periods of the Calabi-Yau threefold
\begin{equation}
  \CA_{\mu} = 2\pi(\md{c}\cdot\md{t}_D + \md{d}\cdot
  \md{t}+ 2\pi\ri m),
\end{equation}
where $\md{t}_D$ are dual periods to $\md{t}$, and it was later
emphasized in \cite{Gu:2023mgf} that with appropriate normalizations,
the coefficients should be integers, so that the action is in fact the
central charge of D-brane bound state with charge
$\mu = (\md{c},\md{d},m)$.

Furthermore, stimulated by a similar connection discovered in complex
Chern-Simons theory
\cite{Garoufalidis:2021osl,Garoufalidis:2020xec,Garoufalidis:2020nut},
it was proposed that these are not just any D-brane bound states, but
stable configurations, and the Stokes constants of the
non-perturbative contributions are their multipliticies, i.e. BPS
invariants or generalized DT invariants
\cite{Marcos:Strings2019,Gu:2021ize,Gu:2023mgf}.  This conjecture was
shown to hold for the resolved conifold in
\cite{Alim:2021mhp,Grassi:2022zuk,Alim:2022oll}, and non-trivial
numerical evidences were provided for more complicated Calabi-Yau
threefolds in \cite{Marino:2024yme,Douaud:2024khu,Douaud:2026qfo}.
Supportive arguments were provided in terms of smoothness of
hypermultiplet moduli space \cite{Alexandrov:2023wdj}, relation to
exact WKB methods \cite{Gu:2023wum}, and close analogy with
wall-crossing fomrulas of Kontsevich and Soibelman for BPS invariants
\cite{Iwaki:2023rst,Douaud:2026qfo}.  The conjecture was later
generalized to refined topological string \cite{Alexandrov:2023wdj},
real topological string \cite{Marino:2023nem}, and topological string
with defects \cite{Grassi:2022zuk,Gu:2024wag}. \footnote{See also
  related developmenet
  \cite{Bridgeland:2016nqw,Bridgeland:2017vbr,Bridgeland:2024ecw}.}

It would be certainly desirable to provide a proof or at least more
extensive evidence to these conjectures.  One obstruction to achieving
this is that except for simple examples 
perturbative free energy can only be computed to a finite
order, either due to theoretical constraints or due to limited
computational resources.  Consequently, only non-perturbative sectors
with relatively small $\CA$ in absolute value can be accessed by the
resurgence theory, and the corresponding Stokes constants calculated.
However, there are usually infinitely many BPS invariants, and
therefore the comparison between the two cannot be complete.

In this paper, we propose to exploit the structure of the moduli space
to help make the comparison between Stokes constants and BPS
invariants.  The moduli space of topological string has rich
structures, including singular points and stability chambers separated
from each other by walls of marginal stability.
Making a loop around each singular point, the periods undergo
monodromy transformations $M_\star$, and they generate the monodromy
group which is a subgroup of the symplectic group of the periods.  It
is known that the non-holomorphic free energies of topological string
are invariant under actions of the monodromy group. The first result
of our paper is that the non-holomorphic non-perturbative
contributions to topological string free energies are not invariant,
but they transform covariantly with respect to the monodromy group.
Furthermore, by exploiting resurgence techniques, we prove that inside
each stability chamber non-holomorphic non-perturbative contributions
are related to each other by local monodromy group induced by singular
points inside the chamber, such that $\CA_\mu$ are related to each
other by $\mu\rightarrow \mu\cdot M_\star$, and the associated Stokes
constants are the same; in other words, non-perturbative sectors form
orbits of the local monodromy group.  Note that in the holomorphic
limit, this was already observed empirically in
\cite{Gu:2023mgf,Marino:2023gxy,Douaud:2026qfo}, but not understood
why.  Furthermore, we demonstrate the power of the local monodromy
invariance of Stokes constants by showing that in some examples of
topological string models assoicated to four dimensional gauge
theories such as Seiberg-Witten theory and 4d massless $N_f=4$ SQCD
that we can produce infinitely many Stokes constants and they recover
the entire BPS spectrum in all stability chambers of the theory.

When we cross the walls of marginal stability and traverse the
stability chambers, the BPS states may merge or decay so that the BPS
invariants would change.  The change of the BPS spectrum is controled
by the Kontsevich-Soibelman wall-crossing formula
\cite{Kontsevich:2008fj}, which states that for each BPS state is 
associated a so-called Kontsevich-Soibelman transformation, the
generators of different KS transformations satisfy certain Lie
algebraic relations, and that a particular product of these
Kontsevich-Soibelman transformations known as the spectrum generator
which depend on the BPS invariants remains the same across the walls.
Similar phenomenon happen to non-perturbative contributions to
topological string free energy.  Following the resugence theory,
different non-perturbative sectors can be represented as singular
points of the so-called Borel transform of the perturbative free
energy.  The singular points may decay or merge across walls of
marginal stability.  For each non-perturbative sector or Borel
singularity is associated a Stokes transformation.  It was shown
\cite{Douaud:2026qfo} that in the holomorphic limit, the generators of
the Stokes transformations when acting on topological string partition
function satisfy the same KS Lie algebraic relations, and it was
argued that the same product of Stokes transformations naturally
remains the same across a stability wall, provided that the Stokes
constants are identified with the BPS invariants. 
A second result of this paper is that we generalize this argument and
show that the generators of the Stokes transformations on the
non-holomorphic partition function of topological string also satisfy
the KS Lie algebraic relations, and it is not a special feature of the
holomorphc limit.

The remainder of the paper is structured as follows.  In
Section~\ref{sec:hae} we give a quick review of various known results,
including perturbative topological string and the holomorphic anomaly
equations, the essential points of resurgence theory which would be
useful for us, and the compact formulas for non-perturbative
contributions to topological string free energy.  In
Section~\ref{sec:mod}, we reformulate the non-holomorphic
non-perturbative contributions in a modular covariant way, and use the
argument of Borel transform to prove that non-perturbative sectors
form orbits of local monodromy groups and that the Stokes constants
are invariant across the orbits.  In Section~\ref{sec:Lie}, we prove
that the generators of Stokes transformations on the non-holomorphic
partition function also satisfy the KS Lie algebraic relation, paving
the way to further identify the spectrum generator and the global
Stokes transformation.  In Sections~\ref{sec:P2}, \ref{sec:SW}, and
\ref{sec:Nf4}, we illustrate with examples of local $\IP^2$ and
Seiberg-Witten theory, as well as the massless $N_f=4$ SQCD the
modular formalism of non-perturbative contributions, and in particular
in the latter two examples, we demonstrate the power of local
monodromy invariance of Stokes constants to reproduce the entire BPS
spectrum.  Finally, we conclude and make some discussions in
Section~\ref{sec:dis}.

\jjj{Things to do: Derive $F^{(\mu)}$ by applying alien derivative on
  master equation of HAE? Convention $\CA = 2\pi t$? }

\section{A bit of review: topological string and resurgence}

\subsection{Perturbative free energy of topological string}
\label{sec:hae}

Consider topological string theory on a local Calabi-Yau threefold
$X$.  For the simplicity of the discussion, we only consider models
with a single modulus, although the discussion can be easily
generalized to multi-moduli models.  If the topological string model
engineers a field theory, it corresponds to a rank one gauge theory.

The mirror of the Calabi-Yau threefold $X$ is a cone over an elliptic
curve $\Sigma$ with possible punctures, and it can brought to the
Weierstrass form with equation
\begin{equation}\label{eq:Weierstrass}
  y^2 = 4x^3 -g_2(z)x - g_3(z),
\end{equation}
equipped with a meromorphic 1-form $\lambda(z)$ which has the property
that $\pd_z \lambda(z) = \rd x/y$ is the unique holomorphic 1-form on
the elliptic curve.  One can choose a polarization for the elliptic
curve, equivalent to choosing a symplectic basis $(\mu_B,\mu_A,\mu_0)$
for the homology group $H_1(\Sigma)$ where $\mu_{A,B}$ are
non-contractible 1-cycles with intersection
numbers 
\begin{equation}
  \vev{\mu_A,\mu_A} = \vev{\mu_B,\mu_B} = 0,\quad
  \vev{\mu_A,\mu_B} = -\vev{\mu_B,\mu_A} = \kappa \in \IN,
\end{equation}
and $\mu_0$ is a puncture on $\Sigma$.
This defines a set of periods
\begin{equation}
  \Pi = \tsp{(t_D,t,t_0)} = \tsp{\left(\int_{\mu_B}\lambda,\int_{\mu_A}\lambda,\int_{\mu_0}\lambda\right)},
\end{equation}
where $t_0$ is simply a constant.  The two non-trivial periods $t_D$
and $t$ are then related by the modulus parameter $\tau$ of the
elliptic curve through
\begin{equation}\label{eq:tau}
  \tau = \frac{\pd t_D}{\pd t}.
\end{equation}
In physics, this choice of polarisation is often called choosing a
frame.

The moduli space $\CM$ of complex structure parametrized by $z$ can be
mapped to the upper half plane $\CH$ of $\tau$ through the $J$-invariant
\begin{equation}\label{eq:Jinv}
  J = \frac{1728g_2^3(z)}{g_2^3(z)-27g_3^2(z)} = \frac{1728
    E_4^3(\tau)}{E_4^3(\tau) - E_6^2(\tau)} = \frac{1}{q} + 744 +
  196884 q + \ldots.
\end{equation}
The moduli space $\CM$ has singular points that are mapped to cusp points of
$\CH$, which are described by zeros of the discriminant
\begin{equation}
  \Delta(z) = g_2^3(z)-27g_3^2(z).
\end{equation}
Circling around a cusp point $\star$, the periods undergo a
non-trivial monodromy
\begin{equation}
  \Pi \rightarrow M_\star \cdot \Pi.
\end{equation}
Near a cusp point, we can choose a polarization where the period $t$
has trivial monodromy around the cusp\footnote{The monodromy of $t$
  can involve the constant period $t_0$ but not the non-trivial period
  $t_D$.}, so that the period $t$ can serve as a proper local flat
coordinate in the neighborhood of the cusp point.  It obeys the
equation \cite{Brandhuber:1998pcc}\footnote{As explained in
  \cite{Huang:2013yta}, the functions $g_i(z)$ are not invariant as
  the Weierstrass form \eqref{eq:Weierstrass} is not changed under the
  scaling
  $g_i\rightarrow c(z)^{2i} g_i, x\rightarrow c(z)^2 x, y\rightarrow
  c(z)^3$ for arbitrary $c(z)$, and this freedom as well as the
  normalization constant $c_t$ is fixed by requiring particular
  leading behavior of $t(z)$.}
\begin{equation}\label{eq:dadz}
  \frac{\rd t}{\rd z} = c_t\sqrt{\frac{g_2(z)}{g_3(z)}\frac{E_6(\tau)}{E_4(\tau)}},
\end{equation}
where $E_{4}(\tau), E_{6}(\tau)$ are the Eisenstein series, and the
cusp point is approached in the limit $\tau\rightarrow \ri\infty$.

The cusp points can be classified by the equation satisfied by the
associated monodromy \cite{Landman:1973opl,Straten:2018cyo} 
\begin{equation}
  (M_\star^k-\md{1})^{p+1} = 0.  
\end{equation}
One has, for instance, a maximal unipotent (MUM) point if $p=2$, a
conifold (C) point if $p=1$, and an orbifold (O) point if $p=0$ and
$k>1$.  The monodromy matrices from all the singular points generate
the monodromy group $\Gamma$, which is a congruent subgroup of
$SL(2,\IZ)$.  Note that the coordinate $z$ is always invariant under
the action of the monodromy group, as after circling around a cusp
point we always come back to the same position on the moduli space.
On the other hand, the flat coordinate $t$ is not modular invariant;
in fact, it is clear from \eqref{eq:dadz} that $(\rd t/\rd z)^2$ is
modular with respect to $\Gamma$ of weight two\footnote{As emphasized
  in \cite{Huang:2011qx}, $(\rd t/\rd z)^2$ is \emph{not} modular with
  respect to the entire $SL(2,\IZ)$.}. 

We can define the perturbative series of free energies of topological
string, 
\begin{equation}
  F(z,\bar{z};g_s) = \sum_{g=0}^\infty g_s^{2g-2} F_g(z,\bar{z}).
\end{equation}
The holomorphic limit of the free energies
\begin{equation}\label{eq:hol-lim}
  \bar{\tau}\rightarrow -\ri\infty, \; F_g(z,\bar{z})\rightarrow \CF_g(z),
\end{equation}
often has enumerative meanings in some frames and is therefore
important.
In this paper, we will nevertheless not restrict ourselves only to the
holomorphic limit, and often consider the non-holomorphic free
energies $F_g$.  The genus zero free energy, i.e.~the prepotential, is
always holomorphic and it is defined through the special geometry
relation \cite{Strominger:1990sgm,Candelas:1990pi} 
\begin{equation}\label{eq:sgr}
  t_D = -\frac{\kappa}{2\pi\ri}\frac{\pd \CF_0}{\pd t}.
\end{equation}
Higher genera free energies can be defined as solutions to the
holomorphic anomaly equations
\cite{Bershadsky:1993ta,Bershadsky:1993cx}.  The genus one case is
special, and for non-compact Calabi-Yau threefolds with a single
modulus, its solution can be taken to be \cite{Huang:2009md}
\begin{equation}\label{eq:F1}
  F_1 = - \log \left(\sqrt{\tau_2}|\Phi(\tau)|^2\right)
\end{equation}
where $\tau_2$ is the imaginary part of $\tau$ and $\Phi(\tau)$ is
some modular form of the monodromy group $\Gamma$ of weight $1/2$.  As
$\tau_2$ is modular of weight $(-1,-1)$, $F_1$ is modular invariant.
The holomorphic limit of the genus one free energy is taken to be
\begin{equation}
    \CF_1 = -\log\Phi(\tau).
\end{equation}
The remaining free energies $F_{g\geq 2}$ are solutions of the
recursive equations \cite{Huang:2009md}
\begin{equation}\label{eq:hae-hE2}
  \frac{\pd}{\pd\wh{E}_2} F_g =
  \frac{\kappa}{24}\left(D_t^2F_{g-1} + \sum_{g'=1}^{g-1}D_tF_{g'}D_tF_{g-g'}\right).
\end{equation}
Here $D_t$ is the covariant derivative on the moduli space and it acts
on the free energies by 
\begin{equation}
  D_t F_g = \pd_t F_g,\quad D_t^2 F_g = (\pd_t - \Gamma^t_{tt})\pd_t F_g,
\end{equation}
with the Christopher symbol,
\begin{equation}
  \Gamma^t_{tt} = \frac{\kappa}{4\pi\tau_2}C_{ttt},
\end{equation}
and the Yukawa coupling is defined through the prepotential
\begin{equation}
  C_{zzz} = \left(\frac{\rd t}{\rd z}\right)^3 C_{ttt} =
  \left(\frac{\rd t}{\rd z}\right)^3 \frac{\rd^3 \CF_0}{\rd t^3}.
\end{equation}
It can be shown that similar to $F_1$ the free energies $F_g$ with
$g\geq 2$ are all modular invariant with respect to the monodromy
group $\Gamma$.  More explicitly, it is convenient to define the
propagator \cite{Huang:2009md,Huang:2011qx,Huang:2013yta}
\begin{equation}\label{eq:SE2}
  S = \frac{\kappa}{12}\left(\frac{\rd t}{\rd z}\right)^{-2}\wh{E}_2(\tau,\bar{\tau})  
\end{equation}
in terms of the famed non-holomorphic Eisenstein series $\wh{E}_2$ related to $E_2$ by
\begin{equation}\label{eq:hE2}
  \wh{E}_2(\tau,\bar{\tau}) = E_2(\tau) - \frac{3}{\pi\tau_2},
\end{equation}
and it is modular invariant with respect to the monodromy group. Then
the free energy $F_g$ can be written as polynomials of the propagator
$S$ \cite{Yamaguchi:2004bt,Alim:2007qj}
\begin{equation}\label{eq:Fg-S}
  F_g = \frac{1}{\Delta^{2g-2}}\sum_{k=0}^{3g-3} S^k P_{g,k}(z).
\end{equation}
The integration constants $P_{g,0}(z)$ are known as holomorphic
ambuiguities as they cannot be fixed by integration and they should be
determined by additional means.  In the case of local Calabi-Yau
threefolds with a single modulus, the holomorphic ambiguities can be
fixed by gap conditions \cite{Ghoshal:1995wm} up to arbitrary genus
$g$ \cite{Haghighat:2008gw}.

\subsection{Review of resurgence theory}
\label{sec:resurgence}


The perturbative series of free energies
\begin{equation}\label{eq:F-pert}
  \wh{F}(g_s) = \sum_{g=2}^\infty F_g g_s^{2g-2},
\end{equation}
defined by and calculated from the holomorphic anomaly equations turn
out to be divergent with zero radius of convergence, as the
coefficients $F_g$ grow in a factorial manner
\cite{Marino:2006hs,Marino:2007te,Marino:2008ya}
\begin{equation}\label{eq:Fg-grow}
  F_g \sim (2g)!\quad\text{for}\quad g\rightarrow\infty,
\end{equation}
which is much faster than exponential.
To make sense of such a divergent power series, one has to introduce a
proper resummation scheme and, as we will shortly see, include
non-perturbative corrections in a suitable way, which can be described
by the mathematical theory of resurgence \cite{Ecalle}.
%
%
We give a short review of the essence of the resurgence theory,
adapted to the perturbative free energy $\wh{F}(g_s)$ which serves as the
principal example.
We will in fact focus on the definition of the alien derivative as
analytic continuation of the Borel transform, which is key to proving
that non-perturbative amplitudes of topological string free energies
are related to each other by modular transformations, and that Stokes
constants are modular invariant under local monodromy actions.
Note that we start in \eqref{eq:F-pert} from $g=2$ so that every term
is modular invariant. \footnote{We also remove $g=1$ as $F_1$ is a bit
  special.}

A divergent power series such as $F(g_s)$ whose coefficients grow like
\eqref{eq:Fg-grow} is called of 1-Gevrey type.  To make sense of such
a divergent series, we first construct an auxiliary power series
called Borel transform
\begin{equation}\label{eq:Btrfs}
  \mr{B}\wh{F}(\zeta) = \sum_{g=2}^\infty \frac{F_g}{(2g-2)!}\zeta^{2g-2}
\end{equation}
which is convergent with a positive radius of convergence.  If the
Borel transform $\mr{B}\wh{F}(\zeta)$ can be analytically continued to
infinity along certain direction with orientation $\theta$ and its
absolute value does not grow too fast \footnote{The Borel transform
  should grow at most by exponential
  $\mr{B}\wh{F}(\zeta) \sim \re^{R\zeta}$, in which case the Borel
  resummation is convergent if $\real \re^{\ri\theta}(1/g_s-R)>0$.},
we can define its Laplace transform along this direction,
\begin{equation}\label{eq:Bsum}
  \mr{S}^{(\theta)}\wh{F}(g_s) = 
  \frac{1}{g_s}\int_0^{\re^{\ri\theta}\infty}\rd \zeta
  \re^{-\zeta/g_s}\mr{B}\wh{F}(\zeta),
\end{equation}
which is convergent if $|\arg g_s - \theta| < \pi/2$, and it is called
the (directed) Borel resummation of the power series.
Formally, the formula of Borel resummation can be obtained by
insertion of $1=\Gamma(2g-1)/(2g-2)!$ in the 1-Gevrey series followed
by exchanging the order of the infinite sum and the integral that
appears in the definition of $\Gamma(2g-1)$,\footnote{In the last
  step, the direction $\theta$ of integration contour can be moved
  away from $\arg g_s$ within the range $|\theta-\arg g_s|<\pi/2$ and
  the integral is still convergent as long as the Borel transform has
  no singularities along this direction.}
\begin{align}
  \wh{F}(g_s)
  &= \sum_{g=2}^\infty \frac{F_g}{(2g-2)!}g_s^{2g-2}\int_{0}^\infty
    \re^{-\zeta}\zeta^{2g-2}\rd \zeta
    ``=" \int_{0}^\infty\rd \zeta \re^{-\zeta} \sum_{g=2}^\infty \frac{F_g}{(2g-2)!}g_s^{2g-2} 
    \zeta^{2g-2}\nn
  &= \frac{1}{g_s}\int_{0}^{\re^{\ri\arg{g_s}}\infty}
    \rd \zeta \re^{-\zeta/g_s}\mr{B}\wh{F}(\zeta).
    \label{eq:BRes-der}
\end{align}
As the power series is divergent, exchanging the infinite sum and the
integration is illegal, but the right hand side can be used to
\emph{define} a resummation of the power series as long as the
integral is convergent.

An important difference of the resummation of a 1-Gevrey power series
from the summation of a convergent power series is that the former
depends on the argument of the summation variable.
%
%
Denote the complex $\zeta$-plane as the Borel plane, and consider an
arbitrary straight ray $d$ in the Borel plane from the origin to
infinity.  The ray is called \emph{active} or a \emph{Stokes ray} if
it passes through a singlar point of the Borel transform.
If the argument of $g_s$ is not far from the direction $\theta$ of an
active ray $d$, one can consider (directed) Borel resummations in two
directions $\theta^{\pm} \equiv \theta \pm \epsilon$, \footnote{The
  condition for the argument of $g_s$ is that
  $\theta+\epsilon-\frac{\pi}{2} < \arg g_s < \theta - \epsilon
  +\frac{\pi}{2}$.}  and their difference, known as the Stokes
discontinuity
\begin{equation}\label{eq:Disc-def}
  \Disc_\theta \wh{F}(g_s) \equiv \mr{S}^{(\theta^+)}\wh{F}(g_s) - \mr{S}^{(\theta^-)}\wh{F}(g_s)
\end{equation}
is non-zero but non-perturbatively small of the order
$\re^{-1/|g_s|}$.  In fact, suppose there is a sequence of singular
points $\{\omega\}$ along the active ray $d$, one can find a
\emph{transseries} of the form
\begin{equation}
  F^{(\omega),+}(g_s) = \re^{-\omega/g_s}\sum_{n=-k_\omega}^\infty a_n^{(\omega),+}g_s^n,
\end{equation}
associated to each singular point $\omega\in d$ so that
\begin{equation}\label{eq:Disc-sum}
  \Disc_\theta \wh{F}(g_s) = \sum_{\omega\in d}\mr{S}^{(\theta^-)}F^{(\omega),+}(g_s),
\end{equation}
where the Borel resummation is defined on the non-negative power
series part of $F^{(\omega),+}(g_s)$, i.e.
\begin{equation}
  \mr{S}^{(\theta)}F^{(\omega),+}(g_s) =
  \re^{-\omega/g_s}\left(\sum_{j=1}^{k_\omega}a_{-j}^{(\omega),+}g_s^{-j}+
    \frac{1}{g_s}\int_0^{\re^{\ri\theta}\infty}\rd
    \zeta \re^{-\zeta/g_s}  \mr{B}\wh{F}^{(\omega),+}(\zeta)\right),
\end{equation}
with
\begin{equation}
  \wh{F}^{(\omega),+}(g_s) = \sum_{n=0}^\infty a_n^{(\omega),+}g_s^n.
\end{equation}

It is clear then that when we move around the origin in the complex
$g_s$ plane, various non-perturbative corrections will come into play,
and each of them is described by a transseries.  All the transseries
form a ring, and we can define the automorphism $\Delta^+_d$ of the
ring of transseries, called \emph{Stokes transformation} associated
to the active ray $d$, such that
\begin{equation}\label{eq:Stk-trfs}
  \mr{S}^{(\theta^+)} = \mr{S}^{(\theta^-)}\Delta^+_d.
\end{equation}
Comparing \eqref{eq:Stk-trfs} with \eqref{eq:Disc-sum}, we have the
relation between Stokes discontinuity and Stokes transformation
\begin{equation}\label{eq:Disc-Del}
  \Disc_\theta = \mr{S}^{(\theta^-)} \left(\Delta^+_d - \md{1}\right).
\end{equation}
It is convenient to define linear components
$\dot{\Delta}^+_{\omega_i}$ of the Stokes transformation
\begin{equation}
  \Delta^+_d = \md{1} + \sum_{\omega\in d}\dot{\Delta}^+_{\omega}
\end{equation}
such that
\begin{equation}
  \dot{\Delta}^+_{\omega}\wh{F}(g_s) = F^{(\omega),+}(g_s).
\end{equation}

There are two major important results in the theory of resurgence
concerning the Stokes transformation \cite{Ecalle,Mitschi:2016fxp}.
The first result is that one can define differential operators
$\dot{\Delta}_\omega$ called \emph{alien derivatives} associated to
each of the singular point $\omega$ such that
\begin{equation}\label{eq:alien}
  \Delta^+_d = \exp \left(\sum_{\omega\in d}\dot{\Delta}_\omega\right).
\end{equation}
This is very useful as the alien derivatives have very nice analytic
properties which give us great computational powers: they can be
treated as genuine derivatives as they satisfy the Leibniz rule
and the chain rule, but they are also alien in the sense that they
live in an orthogonal space to ordinary derivatives with respect to
either the summation variable $g_s$ or additional parameters such as
$z$ or $\bar{z}$ so that they commute with the latter.

The second result is that the Borel transforms of both the Stokes
transformation and the alien derivatives,
\begin{equation}\label{eq:alientwo}
  F^{(\omega),+}(g_s) = \dot{\Delta}_w^+\wh{F}(g_s)\quad
  \text{and}\quad F^{(\omega)}(g_s)\equiv \dot{\Delta}_\omega \wh{F}(g_s),
\end{equation}
are in fact related to the Borel transform of $\wh{F}(g_s)$ via
analytic continuation,
\begin{equation}\label{eq:cont}
  \Cont_\delta \mr{B}\wh{F}(\zeta+\omega) =
  \sum_{j=1}^{k_\omega}\frac{(-1)^{j-1}(j-1)!
    a_{-j}^{(\omega),\delta}}{2\pi\ri\,\zeta^j} +
  \frac{\log(\zeta)}{2\pi\ri}\mr{B}\wh{F}^{(\omega),\delta}(\zeta)
  + \text{reg}.
\end{equation}
Here we make the assumption that the analytic continuation of the
Borel transform $\mr{B}\wh{F}(\zeta)$ is a simply ramified resurgent
function of finite orders \cite{Sauzin:2007resurgent,Douaud:2026qfo},
which means that it has only discrete poles and branch points of
logarithmic type, and no other types of singularities.  This turns out
indeed to be the case for topological string free energies.  In
\eqref{eq:cont}, $\mr{B}\wh{F}^{(\omega),\delta}(\zeta)$ is the Borel
transform of the non-negative power series part of
$F^{(\omega),\delta}(g_s)$ defined by
\begin{equation}
  \mr{B}\wh{F}^{(\omega),\delta}(\zeta) =
  \sum_{n=0}^\infty \frac{a_{n}^{(\omega),\delta}}{n!}\zeta^n
\end{equation}
given that
\begin{equation}
  F^{(\omega),\delta}(g_s) = \sum_{n=-k_\omega}^\infty
  a_n^{(\omega),\delta} g_s^n.
\end{equation}
Here $\delta$ denotes the shortest path of analytic condition from
near the origin to the neighborhood of $\omega$.
If there are additional singular points closer to the origin on the
same active ray $d$ as $\omega$, the path $\delta$ is not unique, and
one needs to specify the way to circumvent the prior singularities.
Two canonical choices are available.  One is to bypass all the prior
singular points by making a semicircle to the right, which is denoted
by $\delta = +$, and it is related to the linear components
$\dot{\Delta}^+_\omega$ of the Stokes transformation.  The other is a
proper weighted sum of all possible paths, in which case, the label
$\delta$ is omitted, and it is related to the alien derivatives
$\dot{\Delta}_\omega$.

By making use of the formula of analytic continuation \eqref{eq:cont},
one can derive a powerful relation between the asymptotic behavior of
the perturbative coefficients $F_g$ as $g\rightarrow \infty$ and the
coefficients of the non-perturbative amplitudes $F^{(\omega),+}(g_s)$,
known as the \emph{resurgence relation},
\begin{equation}\label{eq:rel}
  F_g \sim -\sum_{\omega}\frac{1}{2\pi\ri}\sum_{n=-k_\omega}^\infty
  \frac{a_n^{(\omega),+}}{\omega^{2g-2-n}}\Gamma(2g-2-n).
\end{equation}
This relation can be used recursively to extract the coefficients
$a_m^{(\omega_0),+}$ of the dominant non-perturbative sector with the
smallest $\omega=\omega_0$ in absolute value.  For instance, define
\begin{equation}
  \label{eq:seq-sg}
  s_g^{(j)} \equiv -2\pi\ri \frac{\omega_0^{2g-2-j}}{\Gamma(2g-2-j)}\left(F_g
    + \frac{1}{2\pi\ri}\sum_{n=-k_{\omega_0}}^{j-1}
    \frac{a_n^{(\omega_0),+}}{\omega_0^{2g-2-n}}\Gamma(2g-2-n)\right),
\end{equation}
one has then the asymptotic behavior for $g\rightarrow \infty$
\begin{equation}
  s^{(j)}_g \sim a_j^{(\omega_0),+} + \CO(1/g).
\end{equation}
In addition, coefficients of subdominant non-perturbative sectors can
be extracted after the total contribution of the dominant
non-perturbative sector is removed, which can be achieved by
subtracting \cite{Douaud:2026qfo}
\begin{equation}\label{eq:Igw}
  I^{(\omega)}_g \equiv -\frac{\Gamma(2g-1)}{2\pi\ri}\left(\sum_{n=-k_\omega}^{-1}
    \frac{a_n^{(\omega),+}}{\omega^{2g-2-n}}
    \frac{\Gamma(2g-2-n)}{\Gamma(2g-1)} +\int_0^{\re^{\ri(\arg\omega)_-}\infty} \frac{\mr{B}\wh{F}^{(\omega),+}(\zeta)}{(\zeta+\omega)^{2g-1}}\rd\zeta\right),
\end{equation}
for the dominant sector $\omega=\omega_0$ from $F_g$.
This procedure can be repeated recursively to access information of
progressively higher non-perturbative sectors, until numerical
precision is washed out.

Note that sometimes there is a canonical way to normalize the
non-perturbative amplitudes $F^{(\omega)}(g_s)$, as we will see in the
next section in the case of topological string free energy, so that
there will be a proportionality constant in front on the right hand
side of \eqref{eq:cont} or \eqref{eq:alientwo}, which are called
Stokes constants.  In this section we have absorbed the Stokes
constants in the non-perturbative amplitudes.



\subsection{Non-perturbative amplitudes}
\label{sec:np}


The non-perturbative free energies $F^{(\omega),\delta}(g_s)$ were
calculated in \cite{Gu:2022sqc,Gu:2023mgf} by solving the transseries
solution to the holomorphic anomaly equations
\cite{Couso-Santamaria2016,Couso-Santamaria2015}.  We summarize the
results in the following, and focus on the case of local Calabi-Yau
threefolds with a single modulus.

A singularity of the Borel transform $\mr{B}\wh{F}(\zeta)$ is given by
an integral period \jjj{A new convention!}
\begin{equation}\label{eq:Amu}
  \CA_\mu = 2\pi \mu\cdot \Pi = 2\pi (pt_D + q t + r t_0),
\end{equation}
where
\begin{equation}
  \mu = (p,q,r),
\end{equation}
is the integer charge vector.  Note that $\mu$ depends on the choice
of frame $\Pi$.  We define an associated differential operator
$\ms{D}_\mu$ in the following manner.  We demand that $\ms{D}_\mu$
satisfy the Leibniz rule
\begin{equation}
  \ms{D}_\mu(fg) = \ms{D}_\mu(f) g + f\ms{D}_\mu(g)
\end{equation}
and that its action on free energies or their derivatives be the
following
\begin{equation}\label{eq:SD0}
  \ms{D}_\mu =
  \begin{cases}
    g_s\CA_\mu/\CF_0\quad &\text{acting on } \CF_0,\\
    g_s\pd_z \CA_\mu(S - \CS_{\CA_\mu})\pd_z\quad &\text{else}.
  \end{cases}
\end{equation}
Here $\CS_{\CA_\mu}$ is the holomorphic limit of the propagaor in a
distinguished frame with the polarization
$\wt{\Pi} = (\tilde{t}_D,\tilde{t})$ where $\CA_\mu$ is an $A$-period,
i.e.
\begin{equation}\label{eq:A-dis-pol}
  \CA_\mu = 2\pi (\ell \tilde{t} + r t_0).
\end{equation}
The polarization of the distinguished frame can related to that of a
generic frame \eqref{eq:Amu} with $p\neq 0$ by a modular transformation
\begin{equation}\label{eq:mod-trfs}
\begin{pmatrix}
    t_D \\ t
  \end{pmatrix}\rightarrow
  \begin{pmatrix}
    \tilde{t}_D\\ \tilde{t}
  \end{pmatrix} =
  \begin{pmatrix}
    a & b \\ c & d
  \end{pmatrix}\cdot
  \begin{pmatrix}
    t_D \\ t
  \end{pmatrix},\quad \gamma = \begin{pmatrix}
    a & b \\ c & d
  \end{pmatrix}\in SL(2,\IZ),
\end{equation}
and the coefficients of the charge vectors in the two frames are
related by
\begin{equation}
  p = \ell c,\quad q = \ell d.
\end{equation}

The claim is then that the alien derivative when acting on the
partition function 
\begin{equation}
  Z = \exp F = \exp \sum_{g=0}^\infty g_s^{2g-2}F_g
\end{equation}
is given by \cite{Gu:2022sqc} \jjj{A quick derivation?}
\begin{equation}\label{eq:DelDZ}
  \dot{\Delta}_\mu Z = 
  \frac{\overline{S}_\mu}{2\pi\ri}\left(1+\ms{D}_\mu\right)
  \re^{-\ms{D}_\mu} Z.
\end{equation}
Here the proportionality constant $\overline{S}_\mu$ is called the
Stokes constant, and it is independent of the moduli.  The action of
the alien derivative on the free energy can be calculated using the
fact that the pointed alien derivative is a derivation.  For instance
\begin{align}
  \dot{\Delta}_\mu F =
  &\frac{\overline{S}_\mu}{2\pi\ri} Z^{-1}\left(1+\ms{D}_\mu\right)
    \re^{-\ms{D}_\mu} \exp F \nn=
  &\frac{\overline{S}_\mu}{2\pi\ri} \re^{\re^{-\ms{D}_{\mu}}F-F}
    \left(1+ \ms{D}_\mu\left(\re^{-\ms{D}_\mu}F\right)\right) \equiv
    \frac{\overline{S}_\mu}{2\pi\ri} F^{(\mu)},
    \label{eq:Fmu}
\end{align}
where we have used that $\ms{D}_\mu$ is a derivation and
$\re^{-\ms{D}_\mu}$ is an automorphism.  Using the analytic properties
of $\dot{\Delta}_\mu$, repeated actions
$\dot{\Delta}_{\mu_{n}}\circ \ldots\circ\dot{\Delta}_{\mu_{1}} Z$ or
$\dot{\Delta}_{\mu_{n}}\circ \ldots\circ\dot{\Delta}_{\mu_{1}} F$ can
also be calculated.  For instance \cite{Douaud:2026qfo},
\begin{equation}\label{eq:DelDnZ}
  \dot{\Delta}^n_\mu Z =
  \left(\frac{\overline{S}_\mu}{2\pi\ri}\right)^{n}(1+\SD_\mu)^n
  \re^{-n\SD_\mu} Z,\quad n\geq 1.
\end{equation}
Note that as the alien derivatives annihilate truncated power series,
we naturally have 
\begin{equation}
  \dot{\Delta}_\mu \wh{F} = \dot{\Delta}_\mu F,
\end{equation}
where $\wh{F}$ is defined in \eqref{eq:F-pert}.

An important conjecture concerning the Stokes constants is that they
can be decomposed in terms of more primitive ingredients
\cite{Douaud:2026qfo}
\begin{equation}\label{eq:Sbar}
  \overline{S}_\mu = \sum_{n|\mu} \frac{S_{\mu/n}}{n^2},
\end{equation}
so that the so-called primitive Stokes constants $S_\mu$ are integers
and they coincide with the integer generalized Donaldson-Thomas
invariants $\Omega(\mu)$ with charge $\mu$
\cite{Marcos:Strings2019,Gu:2021ize,Gu:2023mgf,Gu:2023wum,Iwaki:2023rst}.
From the physics point of view, the generalized DT invariants are the
counts of BPS states of either D4-D2-D0 branes in type IIA superstring
on $X$ or D3-D1-D(-1) branes in type IIB supserstring on the mirror
$X^*$.  In the language of IIA superstring, the charges $p,q,r$ refer
to respectively the numbers of D4, D2 and D0 branes.  In the special
case of absence of D4 branes, the generalized DT invariants coincide
with integer genus zero GV invariants
\cite{Joyce:2012atg,Alexandrov:2015qhm},
\begin{equation}
  \Omega(0,q,r) = n_{0,q},
\end{equation}
which also count the D2-D0 BPS states.  These BPS states arise from
wrapping M2 branes in M theory on curve classes in the threefold $X$,
giving rise to the BPS states \cite{Gopakumar:1998ii,Gopakumar:1998jq}
\begin{equation}
  [(\tfrac{1}{2},0)\oplus 2(0,0)]
  \otimes\bigoplus_{j_L,j_R}N_{j_L,j_R}^{q}[(j_L,j_R)].
\end{equation}
After factoring out a universal center-of-mass hypermultiplet, the
integers $N_{j_L,j_R}^{q}$ count BPS states with charge $q$
transforming with spin $(j_L,j_R)$ under the little group
$SU(2)_L\times SU(2)_R$ in 5 dimensions, and are sometimes called BPS
invariants.  The genus zero GV invariant is then
\begin{equation}
  n_{0,q} = \sum_{j_L,j_R} (-1)^{2j_L+2j_R}(2j_L+1)(2j_R+1) N_{j_L,j_R}^q.
\end{equation}
For instance, a hypermultiplet with $(j_L,j_R)=(0,0)$ contributes $1$
to the GV invariant, while a vectormultiplet with
$(j_L,j_R) = (0,\tfrac{1}{2})$ contributes $-2$.  In general, however,
the generalized DT invariants are more broadly defined for BPS states
with arbitrary D4-D2-D0 brane charges.

\section{Modular formulation of non-perturbative amplitudes}
\label{sec:mod}

\subsection{Modular formulation of non-perturbative amplitudes}

Given the modular properties of the free energies $F_g$ reviewed in
Section~\ref{sec:hae}, it is desirable to have a modular formulation
of the non-perturbative free energies as well.  We can first work in
the distinguished frame where $\CA$ is an period with charge vector
\eqref{eq:A-dis-pol}.  The non-holomorphic progator and its
holomorphic limit are respectively, cf.~\eqref{eq:SE2},
\begin{equation}
  S = \frac{\kappa}{12}\left(\frac{\rd \tilde{t}}{\rd
      z}\right)^{-2}\wh{E}_2(\tilde{\tau},\bar{\tilde{\tau}}),\quad
  \CS_{\CA} = \frac{\kappa}{12}\left(\frac{\rd \tilde{t}}{\rd
      z}\right)^{-2}E_2(\tilde{\tau}),
\end{equation}
and thus when not acting on $\CF_0$ the derivative $\SD_\mu$ reads
\begin{equation}
  \SD_\mu = -g_s\frac{\kappa\ell}{2\tau_2}\frac{\pd}{\pd \tilde{t}}.
\end{equation}
The expression in a generic frame can be obtained by a modular
transformation \eqref{eq:mod-trfs},
\begin{equation}\label{eq:SD-else}
  \SD_\mu = -\kappa g_s \frac{p\bar{\tau}+q}{2\tilde{\tau}_2}\pd_t.
\end{equation}
Therefore, eq.~\eqref{eq:SD0} can be written in the simple form
\begin{equation}\label{eq:SD}
  \ms{D}_\mu =
  \begin{cases}
    g_s\CA_\mu/\CF_0\quad &\text{acting on } \CF_0\\
    \displaystyle -\kappa g_s \frac{p\bar{\tau}+q}{2\tilde{\tau}_2}\pd_t,\quad &\text{else}.
  \end{cases}
\end{equation}
This is our first result.  We note that eq.~\eqref{eq:SD-else} can
also be written as
\begin{equation}
  \SD_\mu = -\kappa g_s \frac{\overline{\pd_t \CA_\mu}}{2\tau_2}\pd_t,
\end{equation}
which can be easily generalized to models with more than one modulus.

As a sanity check, we can consider the holomorphic limit of
\eqref{eq:SD}.  Taking the limit $\bar{\tau}\rightarrow -\ri\infty$ so
that $\bar{\tau}/\tau_2\rightarrow -2\ri$, the derivative $\SD_\mu$
becomes
\begin{equation}
  \ms{D}_\mu\rightarrow \CD_\mu =
  \begin{cases}
    g_s\CA_\mu/\CF_0\quad &\text{acting on } \CF_0= F_0\\
    0\quad &\text{else and } p =0,\\
    \ri g_s p \kappa\pd_t \quad &\text{else and } p \neq 0.
  \end{cases}
\end{equation}
In a distinguished frame where $p=0$, we thus have
\begin{equation}
  \re^{-\CD_\mu} \CF = \CF - \CA_\mu/g_s
\end{equation}
and
\begin{equation}
  \CD_\mu\re^{-\CD_\mu} \CF = \CD_\mu (\CF - \CA_\mu/g_s) =  \CA_\mu/g_s
\end{equation}
so that
\begin{equation}\label{eq:DelF-A}
  \dot{\Delta}_\mu\CF = \frac{\overline{S}_\mu}{2\pi\ri}
  \left(1+\CA_\mu/g_s\right)\re^{-\CA_\mu/g_s}.
\end{equation}
which agrees with \cite{Pasquetti2010}.  In a generic frame where
$p\neq 0$, one has that
\begin{align}
  \re^{-\CD_\mu}\CF =
  &g_s^{-2}\sum_{n=0}^\infty
    \frac{(-1)^n}{n!}\CD_\mu^n \CF_0 + \sum_{g=1}^\infty g_s^{2g-2}
    \sum_{n=0}^\infty\frac{(-1)^n}{n!}\CD_\mu^n \CF_g\nn=
  &g_s^{-2}\CF_0 + g_s^{-1}\sum_{n=1}^\infty
    \frac{(-1)^n}{n!}\CD_\mu^{n-1} \CA_\mu
    + \sum_{g=1}^\infty g_s^{2g-2} \exp(-\CD_\mu) \CF_g,
    \label{eq:dDF}
\end{align}
and also
\begin{align}
  \CD_\mu\re^{-\CD_\mu}\CF =
  &g_s^{-1}\CA_\mu + g_s^{-1}\sum_{n=1}^\infty
    \frac{(-1)^n}{n!}\CD_\mu^{n} \CA_\mu
    + \sum_{g=1}^\infty g_s^{2g-2} \CD_\mu\exp(-\CD_\mu) \CF_g.
\end{align}
Define $\CF_{\mu,0},\delta_\mu\CF_0,\CF_\mu$ by
\begin{equation}
  p\kappa\pd_t\CF_{\mu,0} = \CA_\mu,\quad \CF_{\mu,0} = \CF_0 +
  \delta_\mu \CF_0,\quad \CF_\mu = \CF + \delta_\mu \CF_0,
\end{equation}
the above results can be written as
\begin{equation}
  \re^{-\CD_\mu}\CF = g_s^{-2} (\CF_0 -\CF_{\mu,0}) +
  \CF_\mu(t-\ri\kappa pg_s),
\end{equation}
and
\begin{equation}
  \CD_\mu\re^{-\CD_\mu}\CF =p g_s \pd_t \CF_\mu(t-\ri\kappa pg_s).
\end{equation}
Therefore 
\begin{equation}\label{eq:DelF-B}
  \dot{\Delta}_\mu \CF =
  \frac{\overline{S}}{2\pi\ri}\re^{\CF_\mu(t+\kappa pg_s)-\CF_\mu(t)}\left(1+p
    g_s\pd_t\CF_\mu(t-\ri\kappa p g_s)\right),
\end{equation}
which is in agreement with \cite{Gu:2022sqc}.

We notice that while the perturbative free energy $\wh{F}$ is modular
invariant with respect to the monodromy group, the non-perturbative
amplitudes
\begin{equation}
    F^{(\mu)} = \re^{\re^{-\SD_\mu}F-F}\left(1+\SD_\mu(\re^{-\SD_\mu}F)\right)
\end{equation}
are not.  Instead non-perturbative amplitudes in different sectors are
related to each other by modular transformations.  In fact, one can
show that under a modular transformation with respect to the monodromy
group $\Gamma$
\begin{equation}\label{eq:mod-trfs-gen}
    \begin{pmatrix}
        t_D\\t 
    \end{pmatrix}\rightarrow 
    \gamma \cdot
    \begin{pmatrix}
        t_D\\t
    \end{pmatrix},\quad \gamma=
    \begin{pmatrix}
        a & b \\ c & d
    \end{pmatrix}\in \Gamma \subset SL(2,\IZ),
\end{equation}
the non-perturbative amplitude is transformed to
\begin{equation}\label{eq:Fmu-mod}
  F^{(\mu)}\rightarrow F^{(\mu\cdot\gamma)},
\end{equation}
with
\begin{equation}
    \mu\cdot\gamma = (pa+qc,pb+qd,r).
\end{equation}
Indeed, the non-perturbative amplitude can be written as
\begin{equation}
    F^{(\mu)} =  \exp\left( \sum_{k=1}^\infty \frac{(-1)^k}{k!}\SD_\mu^k F\right)\left(1 + \sum_{k=1}^\infty \frac{(-1)^{k-1}}{(k-1)!}\SD_\mu^{k}F\right)
\end{equation}
which consist of $\SD_\mu^kF$ for $k\geq 1$.  At $k=1$
\begin{equation}
    \SD_\mu F = \CA_\mu/g_s +\sum_{g=1}^\infty g_s^{2g-2}\SD_\mu F_g.
\end{equation}
It is easy to check that after a modular transformation
\eqref{eq:mod-trfs-gen}, the component $\CA_\mu$ and the derivative
$\SD_\mu$ become
\begin{equation}\label{eq:AD-mod}
    \CA_\mu \rightarrow \CA_{\mu\cdot\gamma},\quad
    \SD_\mu \rightarrow \SD_{\mu\cdot\gamma},
\end{equation}
while $F_g$ for $g\geq 1$ are invariant, so that
\begin{equation}
  \SD_\mu F \rightarrow \SD_{\mu\cdot\gamma }F.
\end{equation}
Similar arguments can be made for $\SD^k_\mu F$ with $k > 1$.  Thus
the non-perturbative amplitude $F^{(\mu)}$ transforms under the
monodromy group action \eqref{eq:mod-trfs-gen} according to
\eqref{eq:Fmu-mod}.

\subsection{Monodromy property of Stokes constants}
\label{sec:mon}

The discussion of the modular properties of non-perturbative
amplitudes in the previous section paves the way to understand the
modular properties of the Stokes constants.
According to the resurgence theory and in particular \eqref{eq:cont},
the non-perturbative amplitudes $F^{(\mu)}$ are related to the
perturbative series $F$ by the analytic continuation of their
respective Borel transforms
\begin{equation}\label{eq:F-cont}
  \Cont_\delta \mr{B}\wh{F}(\zeta) =
  \frac{\overline{S}_\mu}{2\pi\ri}\left(\sum_{j=1}^{k_\mu}\frac{(-1)^{j-1}(j-1)!
      a_{-j}^{(\mu)}}{2\pi\ri\,(\zeta-\CA_\mu)^j} +
    \frac{\log(\zeta-\CA_\mu)}{2\pi\ri}\mr{B}\wh{F}^{(\mu)}(\zeta-\CA_\mu)
    + \text{reg}\right).
\end{equation}
Here recall that $\mr{B}\wh{F}^{(\mu)}$ is the Borel transform of the
non-negative power part while $a_{-j}^{(\mu)}$ are the coefficients of
the polar part of $F^{(\mu)}$.  Now let us move continuously in the
moduli space $\CM$ (and simultanously in $\overline{\CM}$), circle
around a cusp point, and come back to the original position, and
consider what happens on both sides of the identity. In this process,
we always keep the variable $\zeta$ fixed.

%

On the l.h.s.\ of \eqref{eq:F-cont}, as we have discussed in
Section~\ref{sec:hae}, the components of the perturbative free energy
$\wh{F}(z,\bar{z};g_s)$ and thus its Borel transform is invariant
after such a process.

Let us consider ingredients on the r.h.s.\ of \eqref{eq:F-cont}.
First of all, when $z$ changes continuously, so do the periods, and
all the singularities $\CA_\mu$ should also move continuously in the complex
plane.
If at some point in the process of moving in the moduli space, two of
the three basic periods $t_D,t,t_0$ are aligned in the complex plane,
i.e.~are proportional to each other, some singular points $\CA_\mu$
may collide and produce new singlar points or some singular points may
decay to multiple new ones.  In the moduli space, we can define
\emph{walls of marginal stability} of codimension one by the condition
that two of the basis periods are proportional to each other. These
walls divide the moduli space into disjoint stability chambers. We
require that the loops we make in the moduli space are restricted
inside a single stability chamber and do not cross any of these walls.
If this is the case, trajectories of singular points are well
separated from each other.
%
%
When we are back to the original position, a singularity $\CA_\mu$ is
changed by the monodromy action of the basis periods
\eqref{eq:mod-trfs-gen}.  Equivalently, we can keep the basis periods
unchanged, and the charge vector of the singularity $\CA_\mu$ is
mapped from $\mu$ to $\mu\cdot\gamma$ following \eqref{eq:AD-mod}.

Secondly, the non-perturbative amplitude $F^{(\mu)}$ defined by
\eqref{eq:Fmu} consists of $\CA_\mu$ and $F_g$ for $g\geq 1$ as well
as their $\SD_\mu$ derivatives, all of which have known transformation
properties under the monodromy action following \eqref{eq:AD-mod}.  As
we have just argued in the previous section, the non-perturbative
amplitude after transformation is simply $F^{(\mu\cdot\gamma)}$ and we
denote its coefficients by $a_n^{(\mu\cdot\gamma)}$.

Finally, the Stokes constant $\overline{S}_\mu$ associated to the
singularity point $\CA_\mu$ remains the same as it is independent of
the moduli by definition.

Put together, we conclude that if we move around some cusp point
inside a stability chamber in the moduli space, without crossing any
wall of marginal stability, we induce a local monodromy action
$\gamma\subset\Gamma$ on both sides of the equation of analytic
continuation \eqref{eq:F-cont} and find
\begin{align}
  &\Cont_\delta \mr{B}\wh{F}(\zeta) \nn=
  &\frac{\overline{S}_\mu}{2\pi\ri}\left(\sum_{j=1}^{k_\mu}\frac{(-1)^{j-1}(j-1)!
      a_{-j}^{(\mu\cdot\gamma)}}{2\pi\ri\,(\zeta-\CA_{\mu\cdot\gamma})^j} +
    \frac{\log(\zeta-\CA_{\mu\cdot\gamma})}{2\pi\ri}
    \mr{B}\wh{F}^{(\mu\cdot\gamma)}(\zeta-\CA_{\mu\cdot\gamma})
    + \text{reg}\right).
\end{align}
This indicates that if there is a non-perturbative sector labeled by
the charge vector $\mu$ with the Stokes constant $\overline{S}_\mu$
and the non-perturbative amplitude $F^{(\mu)}$, there are also
non-perturbatives sectors labeled by the charge vector
$\mu\cdot\gamma$ with the same Stokes constant $\overline{S}_\mu$ and
the non-perturbative amplitude $F^{(\gamma\cdot\mu)}$ for any such
monodromy action $\gamma$.

Let us emphasize that it is crucial that we stay in the same stability
chamber of the moduli space.  Even though the non-perturbative
amplitude $F^{(\mu)}$ transforms in a universal way under any
monodromy action in the monodromy group, the Stokes constants
$\overline{S}_\mu$ and thus also the primitive ones $S_\mu$ are only
invariant for local monodromy actions induced by loops within a single
stability chamber. If a loop traverses different stability chambers,
the number of singularities of the Borel transform may not stay the
same, and the Stokes constants may change as well.  This is exactly in
parallel to the case of wall-crossing behavior of BPS invariants of
the Calabi-Yau threefold.

The invariance of Stokes constants under local monodromy action has
already been observed empirically in \cite{Gu:2023mgf,Marino:2023gxy},
both near the MUM point and near the orbifold point, and it was also
observed that it is not true near the conifold point, as the latter is
passed through by a wall of marignal stability.
Here we give an argument for this observation.  In the case of local
Calabi-Yau threefolds, the local monodromy invariance is not
particularly powerful.  However, as we will see in examples, for
Calabi-Yau threefolds associated to four dimensional gauge theories,
the local monodromy invariance can be used to generate a large subset
if not all of the Stokes constants, which reproduce the entire
spectrum of BPS invariants of the gauge theory.

\section{Lie algebraic relations for non-perturbative amplitudes}
\label{sec:Lie}

As another piece of evidence that the (primitive) Stokes constants
coincide with generalized DT invariant, one can consider the
following.  Recall that alien derivatives act on the partition
function by, c.f.~\eqref{eq:DelDZ}
\begin{equation}
  \dot{\Delta}_\mu Z = \overline{S}_\mu \CD_\mu Z
\end{equation}
with
\begin{equation}
  \CD_\mu = \frac{1}{2\pi\ri}(1+\SD_\mu)\re^{-\SD_\mu}.
\end{equation}
It was shown in \cite{Douaud:2026qfo} that in the holomorphic limit,
the operators $\CD_\mu$ form a representation of the
Kontsevich-Soibelman Lie algebra, a key ingredient in the
Kontsevich-Soibelman wall-crossing formula for BPS invariants
\cite{Kontsevich:2008fj}, as they satisfy the commutation relations
\begin{equation}\label{eq:Lie-KS}
  [\CD_{\mu_1},\CD_{\mu_2}] =
  (-1)^{\vev{\mu_1,\mu_2}}\vev{\mu_1,\mu_2}\CD_{\mu_1+\mu_2},
\end{equation}
which is the same as the generators of the KS Lie algebra.  As a
consequence, if we define the sectorial Stokes transformation
\begin{equation}
  \mf{S}_V = \prod_{d:\omega\in V}^{\curvearrowright}\Delta_d^+
\end{equation}
from one non-active ray $d_1$ counter-clockwise to another $d_2$ in
the Borel plane, where $V$ is the sector bounded by the two rays, and
the arrow above the product indicates that the product is taken in
clockwise order, it can be identified with the Kontsevich-Soibelman
spectrum generator for BPS invariants whose central charges are also
in the sector $V$, if the primitive Stokes constants $\CS_\gamma$ are
the same as the generalized DT invariants $\Omega(\gamma)$.  Note that
this is quite natural as both the sectorial Stokes transformation
$\mf{S}_V$ and the KS spectrum generator has the wall-crossing
property: upon variation of the moduli of the model, it remains the
same as long as no central charge $Z(\gamma)$ of a BPS state, or
equivalently no singularity $\omega_\gamma$ of the Borel transform,
enters or exits the sector $V$.

In this section, we generalize \cite{Douaud:2026qfo} and argue that at
least for local Calabi-Yau threefolds with a single modulus, the
commutation relations \eqref{eq:Lie-KS} hold true even if we are away
from the holomorphic limit.  We believe our argument can be
generalized to multi-moduli models and even compact Calabi-Yau
threefolds.


Let us first show that when acting on the partition function $Z$, we
have that
\begin{equation}\label{eq:D-com}
  [\SD_{\mu_1},\SD_{\mu_2}] = 
  2\pi\ri\vev{\mu_1,\mu_2}
\end{equation}
where
\begin{equation}
  \vev{\mu_1,\mu_2} = (p_1 q_2 - p_2 q_1)\kappa
\end{equation}
is a symplectic pairing of charge vectors.  For this purpose, we note
that
\begin{equation}
  \SD_{\mu} Z = Z(\SD_\mu F) = Z\left(g_s^{-1}\CA_\mu + \SD_\mu\wt{F}\right),
\end{equation}
where
\begin{equation}
  \wt{F} = F - g_s^{-2} \CF_0.
\end{equation}
Then
\begin{align}
  \SD_{\mu_1}\SD_{\mu_2} Z =
  &\SD_{\mu_1}\left(Z\left(g_s^{-1}\CA_{\mu_2}+\SD_{\mu_2}\wt{F}\right)\right)\nn=
  &Z\left(g_s^{-1}\CA_{\mu_1}+\SD_{\mu_1}\wt{F}\right)
    \left(g_s^{-1}\CA_{\mu_2}+\SD_{\mu_2}\wt{F}\right) +
    Z\left(g_s^{-1}\SD_{\mu_1}\CA_{\mu_2} + \SD_{\mu_1}\SD_{\mu_2}\wt{F}\right),
\end{align}
thus
\begin{equation}\label{eq:DcomZ}
  [\SD_{\mu_1},\SD_{\mu_2}]Z =
  Z\left(g_s^{-1}(\SD_{\mu_1}\CA_{\mu_2}-\SD_{\mu_2}\CA_{\mu_1})+
  [\SD_{\mu_1},\SD_{\mu_2}] \wt{F}
  \right).
\end{equation}
On the one hand
\begin{align}
  \SD_{\mu_1}\CA_{\mu_2} =
  &-\kappa g_s\frac{p_1\bar{\tau}+q_1}{2\tau_2}\pd_t(2\pi(p_2t_D+q_2t+r_2A_0)) =
    - 2\pi\kappa\frac{g_s}{2\tau_2}(p_1\bar{\tau}+q_1)(p_2\tau+q_2)\nn=
  &- 2\pi\kappa\frac{g_s}{2\tau_2}(p_1p_2|\tau|^2+q_1p_2\tau+p_1q_2\bar{\tau}+q_1q_2),
\end{align}
and thus
\begin{equation}\label{eq:DA-com}
  \SD_{\mu_1}\CA_{\mu_2}-\SD_{\mu_2}\CA_{\mu_1} = - 2\pi\kappa
  \frac{g_s}{2\tau_2}
  (q_1p_2-p_1q_2)(\tau-\bar{\tau}) = 2\pi\ri
  g_s\vev{\mu_1,\mu_2}.
\end{equation}
On the other hand,
\begin{align}
  \SD_{\mu_1}\SD_{\mu_2}\wt{F} =
  &g_s^2\frac{p_1\bar{\tau}+q_1}{2\tau_2}\pd_t
    \left(\frac{p_2\bar{\tau}+q_2}{2\tau_2}\pd_t\wt{F}\right)\nn=
  &g_s^2\left(\frac{p_1\bar{\tau}+q_1}{2\tau_2}\right)
    \left(\frac{p_2\bar{\tau}+q_2}{2\tau_2}\right)\pd_t^2\wt{F}+
    \ri g_s^2\frac{(p_1\bar{\tau}+q_1)(p_2\bar{\tau}+q_2)}{(2\tau_2)^3}C_{ttt}\pd_t\wt{F}
\end{align}
where we have used that
\begin{equation}
  \pd_t\left(\frac{p_2\bar{\tau}+q_2}{\tau-\bar{\tau}}\right) =
  -\frac{p_2\bar{\tau}+q_2}{(\tau-\bar{\tau})^2}\pd_t\tau=
  -\frac{p_2\bar{\tau}+q_2}{(\tau-\bar{\tau})^2} C_{ttt}
\end{equation}
and it is clear that
\begin{equation}\label{eq:DF-com}
  [\SD_{\mu_1},\SD_{\mu_2}]\wt{F}  = 0.
\end{equation}
Putting \eqref{eq:DcomZ}, \eqref{eq:DA-com} and \eqref{eq:DF-com} 
together, we get \eqref{eq:D-com}.

Once \eqref{eq:D-com} is established, it is easy to prove that the
operators \jjj{Normalization by $\kappa$?}
\begin{equation}
  \CD_{\mu} := -\frac{1}{2\pi\ri}(1+\SD_\mu)\re^{-\SD_\mu}
\end{equation}
satisfy the Lie algebraic relations \eqref{eq:Lie-KS}.  To see this,
we first calculate,
\begin{align}
  [\re^{-\SD_{\mu_1}},\SD_{\mu_2}] =
  &\sum_{n=0}^\infty\frac{(-1)^n}{n!}[\SD_{\mu_1}^n,\SD_{\mu_2}] =
    - 2\pi\ri\vev{\mu_1,\mu_2}
    \sum_{n=1}^\infty\frac{(-1)^{n-1}}{(n-1)!}
    \SD_{\mu_1}^{n-1} \nn =
  &- 2\pi\ri\vev{\mu_1,\mu_2} \re^{-\SD_{\mu_1}}.
\end{align}
Then we have
\begin{align}
  \CD_{\mu_1} \CD_{\mu_2} =
  &\left(\frac{1}{2\pi\ri}\right)^2
    (1+D_{\mu_1})\re^{-D_{\mu_1}}(1+D_{\mu_2})\re^{-D_{\mu_2}}\nn =
  &\left(\frac{1}{2\pi\ri}\right)^2
    (1+D_{\mu_1})\left(\re^{-D_{\mu_1}}\re^{-D_{\mu_2}}
    +[\re^{-D_{\mu_1}},D_{\mu_2}]\re^{-D_{\mu_2}}
    +D_{\mu_2}\re^{-D_{\mu_1}}\re^{-D_{\mu_2}}
    \right)\nn=
  &\left(\frac{1}{2\pi\ri}\right)^2(1+D_{\mu_1})(1- 2\pi\ri\vev{\mu_1,\mu_2}+D_{\mu_2})
    \re^{-D_{\mu_1}-D_{\mu_2}+\pi\ri\vev{\mu_1,\mu_2}} \nn=
  &(-1)^{\vev{{\mu_1},\mu_2}}\left(\frac{1}{2\pi\ri}\right)^2
    (1+D_{\mu_1}+D_{\mu_2}+D_{\mu_1}
    D_{\mu_2}- 2\pi\ri\vev{\mu_1,\mu_2}(1+D_{\mu_1}))
    \re^{-D_{\mu_1}-D_{\mu_2}},
\end{align}
and therefore
\begin{align}
  [\CD_{\mu_1},\CD_{\mu_2}] =
  &(-1)^{\vev{\mu_1,\mu_2}}\left(\frac{1}{2\pi\ri}\right)^2
    \left([D_{\mu_1},D_{\mu_2}]- 2\pi\ri\vev{\mu_1,\mu_2}(2+D_{\mu_1}+D_{\mu_2})\right)
    \re^{-D_{\mu_1}-D_{\mu_2}}\nn=
  &-\frac{(-1)^{\vev{\mu_1,\mu_2}}
    \vev{\mu_1,\mu_2}}{2\pi\ri}\left(1+D_{\mu_1}+D_{\mu_2}
    \right)\re^{-D_{\mu_1}-D_{\mu_2}}\nn=
  &(-1)^{\vev{\mu_1,\mu_2}}\vev{\mu_1,\mu_2}\CD_{\mu_1+\mu_2},
\end{align}
which is exactly \eqref{eq:Lie-KS}.


\section{Example: Local $\IP^2$}
\label{sec:P2}

We first consider the quintessential example of topological string
theory on local $\IP^2$, which has been studied extensively in the
early days of local mirror symmetry
\cite{Chiang:1999tz,Klemm:1999gm,Haghighat:2008gw}.  This theory can
be described by the mirror curve
\begin{equation}
  \Sigma:\quad \re^x + \re^y + \re^{-x-y} + u = 0,
\end{equation}
with the Batyrev coordinate $z = 1/u^3$ and the canonical 1-form
$\lambda = y \rd x$.  It can be put in the Weierstrass form with
\cite{Huang:2011qx}
\begin{equation}
  \begin{aligned}
    &g_2(z) = 27z^4(1+24z),\\
    &g_3(z) = 27z^6(1+36z+216z^2).
  \end{aligned}
\end{equation}
The basis periods of the curve are \jjj{Check convention}
\begin{equation}
  \begin{aligned}
    t(z) =
    &-\log(z) + 6z
      {}_4F_3(1,1,\tfrac{4}{3},\tfrac{5}{3};2,2,2;-27z),\\
    t_D(z) =
    &\frac{3\ri}{2\pi}\left(\frac{G^{3,2}_{3,3}\left(-27z\big| \begin{smallmatrix}\frac{1}{3},\frac{2}{3},1\\0,0,0\end{smallmatrix}\right)}{2\sqrt{3}\pi}-\frac{4\pi^2}{9}\right),
  \end{aligned}
\end{equation}
and together with the constant period related to the puncture on the
mirror curve they are denoted collectively as
\begin{equation}
  \Pi = \tsp{(t_D,t,2\pi\ri)}.
\end{equation}
A BPS state with electromagnetic charge $\mu= (p,q,r)$ has the central
charge
\begin{equation}
  Z_\mu(z) = \mu\cdot \Pi = p t_D(z) + q t(z) + r 2\pi\ri.
\end{equation}

The moduli space of the local $\IP^2$ has three singular points,
located at $z = 0, -1/27, \infty$ respectively.  Going around a
singular point, the periods undergo monodromy,
\begin{equation}
  \Pi \rightarrow M_\star\cdot \Pi
\end{equation}
and for the three singular points, the monodromy matrices are
respectively, 
\begin{equation}
  M_0 =
  \begin{pmatrix}
    1 & 1 & 1\\
    0 & 1 & 1\\
    0 & 0 & 1
  \end{pmatrix},
  \quad
  M_{-1/27} =
  \begin{pmatrix}
    -2 & 3 & 2\\
    -3 & 4 & 2\\
    0 & 0 & 1
  \end{pmatrix},
  \quad
  M_{\infty} =
  \begin{pmatrix}
    4 & -7 & -2\\
    3 & -5 & -2\\
    0 & 0 & 1
  \end{pmatrix}
\end{equation}
which generate the monodromy group $\Gamma(3)$.

The resurgence properties of the free energies of local $\IP^2$ have
been discussed extensively in the literature
\cite{Couso-Santamaria2015,Gu:2022sqc,Douaud:2026qfo,Couso-Santamaria:2016vwq},
both in the holomorphic limit and without.  It has been found that the
Borel singularities are located at
\begin{equation}
  \CA_\mu = 2\pi Z_\mu,
\end{equation}
which correspond conjecturally to stable D-brane bound states, and
that there has been much evidence about the relationship between
Stokes constants and generalized DT invariants, in particular in the
holomorphic limit.  In this section, we focus on verifying the modular
expression of the non-holomorphic instanton amplitude
\eqref{eq:Fmu},\eqref{eq:SD} near the Large Radius point $z = 0$.

\begin{figure}
  \centering%
  \subfloat[$z=-10^{-2},\tau_2=200$]
  {\includegraphics[height=5cm]{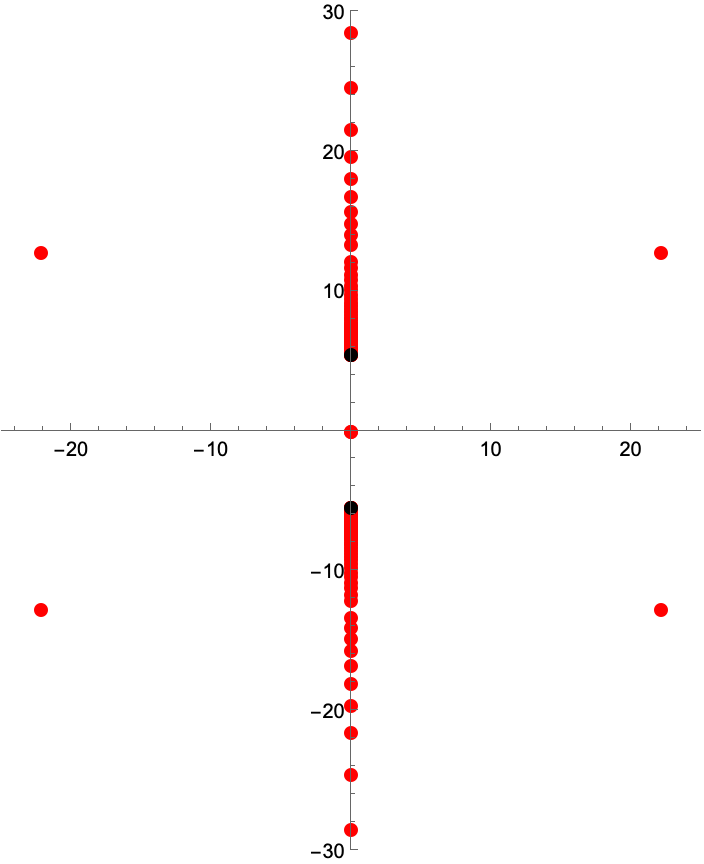}\label{fig:P2zL-brl}}
  \hspace{10ex} \subfloat[$z=10^{-6},\tau_2=200$]
  {\includegraphics[height=5cm]{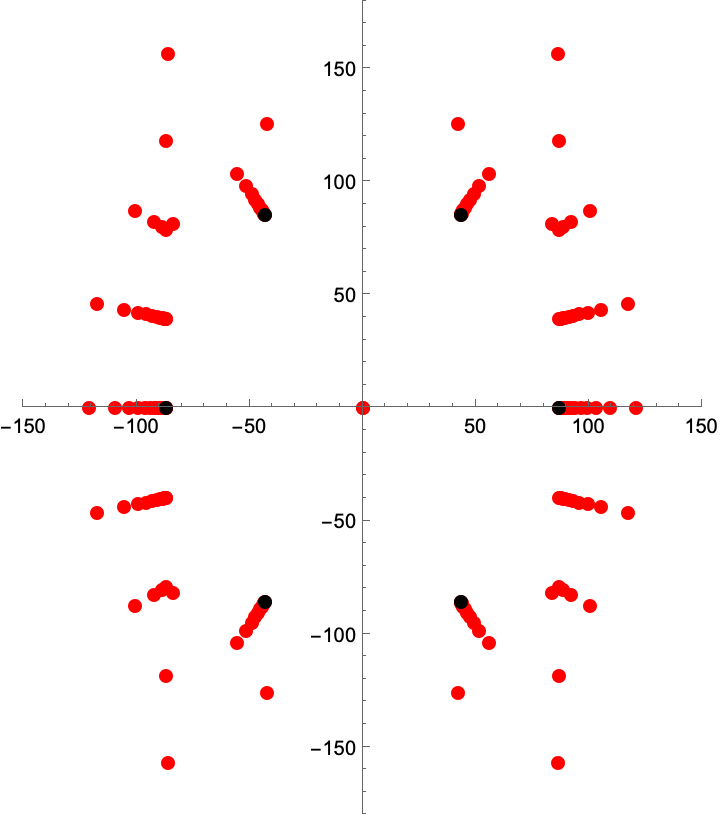}\label{fig:P2zS-brl}}
  \caption{Borel plane of non-holomorphic free energies of local
    $\IP^2$ near the Large Radius point.  For moderately small $z$
    (a), the dominant singularities are $\CA_{(\pm 1,0,0)}$, the two
    black dots on the imaginary axis.  For extremely small $z$ (b),
    the dominant singularities are $\CA_{(0,\pm 1,0)}$, the two black
    dots on the real axis.  The other singular points in the vertical
    towers are $\CA_{(0,\pm 1,n)}$ for $n=\pm 1,\pm 2,\ldots$.  The
    remaining four singular points, which are also marked by black
    dots are $\CA_{(\pm 1,0,0)}$ and $\CA_{(\pm 1,\mp 1,0)}$.}
  \label{fig:P2brl}
\end{figure}

\begin{figure}
  \centering%
  \subfloat[$s^{(-1)}_g
  $]%
  {\includegraphics[height=3cm]{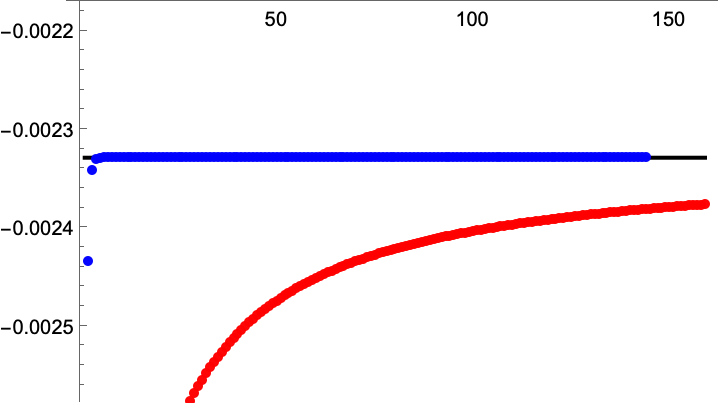}\label{fig:P2zL-b0}}
  \hspace{8ex}
  \subfloat[$s^{(0)}_g
  $]%
  {\includegraphics[height=3cm]{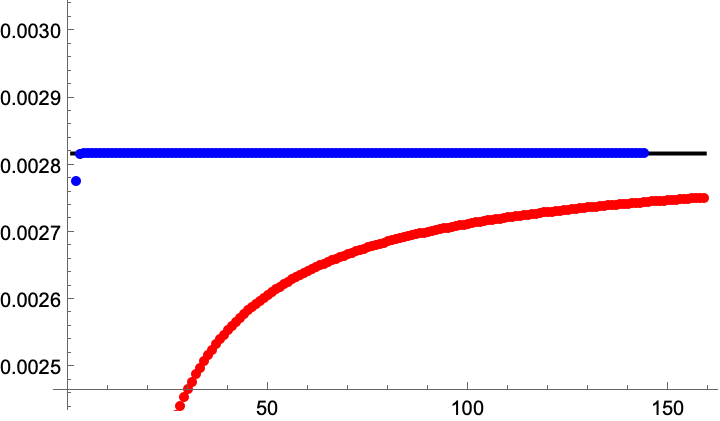}\label{fig:P2zL-b1}}
  \caption{The sequences $s^{(j=-1,0)}_g$ for local $\IP^2$ with
    $z = -10^{-2}$ and $\tau_2 = 200$.  Red dots are numerical values,
    blue dots are results after Richardson transform, and black line is
    the expected asymptotic value, the coefficient $a_j^{(\CA_\mu),+}$
    of non-perturbative amplitude for $\mu=(\pm 1,0,0)$ and $j=-1,0$,
    multiplied with Stokes constant $S_{(\pm 1,0,0)}=1$.}
  \label{fig:P2instcoefs-zM}
\end{figure}

We consider two examples. At $z = -1/100$ with $\tau_2 = 200$, as
shown in Fig.~\ref{fig:P2zL-brl}, the closest Borel singularities are
$\CA_{(\pm 1,0,0)}$.  The coefficients of the associated modular
instanton amplitude can be checked to agree with
\eqref{eq:Fmu},\eqref{eq:SD} by using the resurence relation
\eqref{eq:rel}\footnote{As Borel singularities appear in pairs, the
  resurgence relation should be multiplied by a factor of two.}  and
the auxiliary sequenes \eqref{eq:seq-sg}, as shown in
Fig.~\ref{fig:P2instcoefs-zM}, and one also finds that the Stokes
constant is
\begin{equation}
  S_{(\pm 1,0,0)} = 1,
\end{equation}
which is the same as in the holomorphic limit \cite{Gu:2022sqc}.
Alternatively, at $z = 10^{-6}$ also with $\tau_2 = 200$, the closest
Borel singularities are $\CA_{(0,\pm 1, 0)}$, as shown in
Fig.~\ref{fig:P2zS-brl}.  The associated coefficients of modular
instanton amplitude can again be checked to agree with expectation,
and the Stokes constant is
\begin{equation}\label{eq:S10-P2}
  S_{(0,\pm 1,0)} = 3, 
\end{equation}
again the same as in the holomorphic limit \cite{Gu:2022sqc}.  Both
Stokes constants agree with the corresponding DT invariants of local
$\IP^2$, and $\Omega(0,\pm 1,0)=S_{(0,\pm 1,0)}$ is in fact nothing
else but the first GV invariant of local $\IP^2$.

\begin{figure}
  \centering%
  \subfloat[$s^{(-1)}_g
  $]%
  {\includegraphics[height=3cm]{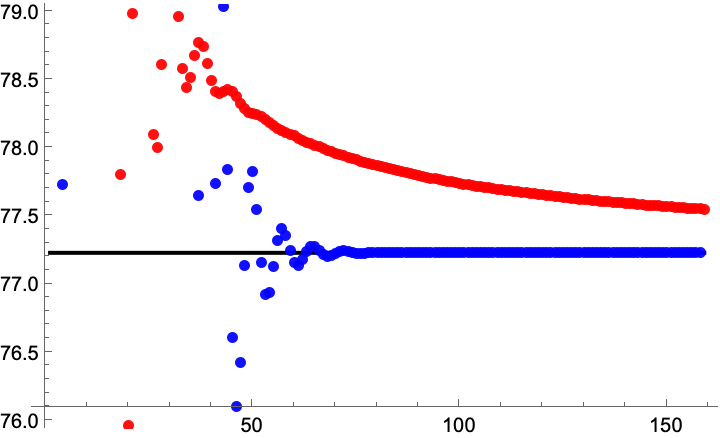}\label{fig:P2zS-b0}}
  \hspace{8ex}
  \subfloat[$s^{(0)}_g
  $]%
  {\includegraphics[height=3cm]{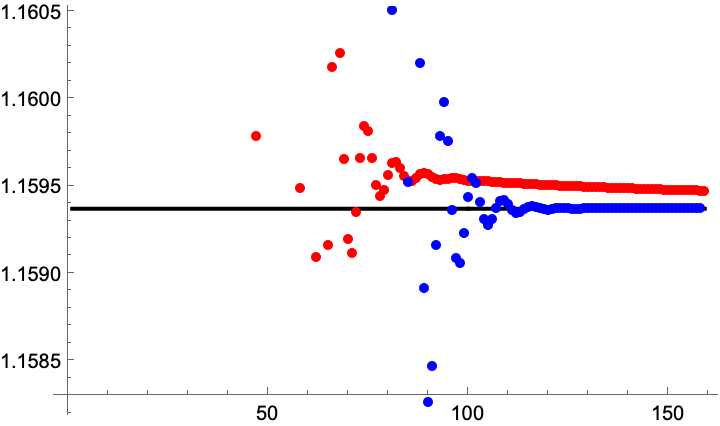}\label{fig:P2zS-b1}}
  \caption{The sequences $s^{(j=-1,0)}_g$ for local $\IP^2$ with
    $z = 10^{-6}$ and $\tau_2 = 200$.  Red dots are numerical values,
    blue dots are results after Richardson transform, and black line is
    the expected asymptotic value, the coefficient $a_j^{(\CA_\mu),+}$
    of non-perturbative amplitude for $\mu=(0,\pm 1,0)$ and $j=-1,0$,
    multiplied with Stokes constant $S_{(0,\pm 1,0)}=3$.}
  \label{fig:P2instcoefs-zS}
\end{figure}

Near the Large Radius point, the only loop path with non-trivial
monodromy action without crossing any wall of marginal stability is
the one around the Large Radius point $z = 0$.  The induced monodromy
matrix is $M_0$, which generates a subgroup $\IZ$ of the monodromy
group $\Gamma(3)$.  By applying the monodromy repeatedly, one can
induce from \eqref{eq:S10-P2} that all the Borel singularities in the
two vertical towers are in the same monodromy orbit of the subgroup
$\IZ$, and they share the same Stokes constant
\begin{equation}
  S_{(0,\pm 1,n)} = 3,
\end{equation}
all in agreement with the first GV invariant of local $\IP^2$.

Similarly, near the orbifold point $z=\infty$, the local monodromy
action by $M_\infty$ generates a subgroup $\IZ_3$ of the monodromy
group $\Gamma(3)$ and it has been shown \cite{Marino:2023gxy} in the
holomorphic limit that Borel singularities form orbits of the subgroup
$\IZ_3$, and the singularities in the same orbit share the same Stokes
constant. 

In general, however, in the case of topological string on local
Calabi-Yau threefolds, the monodromy invariance is not very powerful,
and can only be used to generate a small subset of all Stokes
constants.  The situation is different in topological string model
associated to 4d supersymmetric field theories, where a large subset
if not all of the Stokes constants can be generate by monodromy
invariance from a small set of data, as we will see in the following
examples.

\section{Example: 4d $SU(2)$ Seiberg-Witten theory}
\label{sec:SW}

\subsection{BPS spectrum and perturbative free energy}

We consider the four dimensional $\CN=2$ Super-Yang-Mills theory with
gauge group $SU(2)$ \cite{Seiberg:1994rs}.  This theory  
is described by the Seiberg-Witten curve
\begin{equation}\label{eq:SW-curve}
  \Sigma:\quad  y^2 = 2\Lambda^2\cosh x + 2u
\end{equation}
equipped with the canonical 1-form $\lambda = y\rd x$.  The parameter
$\Lambda$ is the dynamically generated scale, which we will set to 1.
The parameter $u$ is the Coulomb modulus, which also parameterizes the
complex structure of the curve $\Sigma$.
The SW curve can be put in the Weierstrass form with
\cite{Huang:2011qx}
\begin{equation}
  \begin{aligned}
    g_2(u) =
    &\frac{4}{3}u^2-1,\\
    g_3(u) =
    &-\frac{1}{27}(8u^3-9u).
  \end{aligned}
\end{equation}
The basis periods of the curve are \footnote{In line with the
  convention of this paper, we use the notation $t$ to refer to the
  periods, instead of $a$ as common in the literature of gauge
  theory.}
\begin{equation}\label{eq:periods}
  \begin{aligned}
    t(u) =
    &\frac{\sqrt{2}}{\pi}\sqrt{u+1}
      E\left(\frac{2}{u+1}\right),\\
    t_D(u) =
      &\frac{2\ri}{\pi}
        \left((1+u)K\left(\frac{1-u}{2}\right)-2E\left(\frac{1-u}{2}\right)\right)
        ,
  \end{aligned}
\end{equation}
and we denote them
collectively as
\begin{equation}
  \Pi = \tsp{(t_D,t)}.
\end{equation}
We follow the convention of \cite{Huang:2009md} instead of
\cite{Ferrari:1996sv} to conform to the original charge convention
\cite{Seiberg:1994rs}.  Note that in the case of four dimensional
gauge theory there is no constant period.  A BPS state with
electromagnetic charge $\mu=(p,q)$ has then the central charge
\begin{equation}\label{eq:Z}
  Z_\mu(u) = \mu\cdot\Pi = p \, t_D(u) + q \, t(u).
\end{equation}

The Coulomb moduli space of the SYM theory has three singular points:
the point at infinity \cite{Seiberg:1994rs}
\begin{equation}
  u = \infty 
\end{equation}
which corresponds to the weak coupling limit, and additional two points
\begin{equation}
  u = 1, \quad u = -1,
\end{equation}
where the monopole of charges $\mu = \pm (1,0)$ and the dyon of charges
$\mu = \pm (1,2)$ become respectively massless.  Going around a singular
point in the moduli space, the periods undergo monodromy
\begin{equation}
  \Pi \rightarrow M_\star\cdot \Pi,
\end{equation}
and for the three singular points, the monodromy matrices are
respectively
\begin{equation}\label{eq:M-SW}
  M_1 =
  \begin{pmatrix}
    1 & 0 \\ -1 & 1
  \end{pmatrix},
  \quad M_{-1} =
  \begin{pmatrix}
    -1 & 4\\-1 & 3
  \end{pmatrix},
  \quad M_\infty =
  \begin{pmatrix}
    -1 & 4\\ 0 &-1
  \end{pmatrix}.
\end{equation}
They satisfy the constraint that
\begin{equation}
  M_1\cdot M_{-1} = M_\infty,
\end{equation}
and together generate the monodromy group $\Gamma(2)$.

\begin{figure}
  \centering
  \includegraphics[height=5cm]{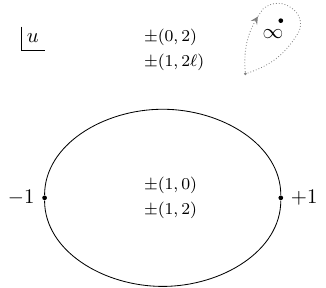}
  \caption{The cusp points and the wall of marginal stability in the
    moduli space of the Seiberg-Witten theory.}
  \label{fig:SW-wall}
\end{figure}

The BPS spectrum of the SYM theory is well known
\cite{Seiberg:1994rs,Ferrari:1996sv}.  There is a wall of marginal
stability defined by
\begin{equation}
  \imag \left(\frac{t_D}{t}\right) = 0
\end{equation}
as shown in Fig.~\ref{fig:SW-wall}, passing through the two singular
points $u=\pm 1$, and it divides the moduli space into the strong
coupling regime inside the wall with small $u$ and the weak coupling
regime outside the wall with large $u$.  In the strong coupling
regime, there are four BPS states of charges
$\mu = \pm (1,0), \pm(1,2)$.  They are hypermultiplets of the
monopole, the dyon, and their anti-particles.  The associated DT
invariants are \footnote{Anti-particles with opposite charges have the
  same multiplicities.}
\begin{equation}\label{eq:OmegaSC-SW}
  \Omega(1,0) =  \Omega(1,2) = 1.
\end{equation}
In the weak coupling regime, the BPS spectrum includes the
vectormultiplets with charges $\mu = \pm (0,2)$ where $W^{\pm}$ bosons
lie, and infinitely many hypermultiplets of dyons with charges
$\mu = \pm (1,2\ell)$ for $\ell \in \IZ$.
The associated DT invariants are
\begin{equation}\label{eq:OmegaWC-SW}
  \Omega(0,2) = -2,\quad \Omega(1,2\ell)  = 1.
\end{equation}

When the 4d gauge theory is coupled to the self-dual limit of the
Omega background, it is equivalent to a topological string model.  It
can be viewed as certain four-dimensional limit of topological string
theory on a non-compact Calabi-Yau threefold \cite{Katz:1996fh}, or
more directly one can define its free energies by solving holomorphic
anomaly equations \cite{Bershadsky:1993cx} on the curve
\eqref{eq:SW-curve}, as first pointed out by \cite{Huang:2009md}.
This was later also carried in \cite{Marino:2024yme} following a more
convenient formulation in \cite{Gu:2022sqc}.
We summarize the results in the following.

We first take the electric frame where we choose $t$ to be the local
flat coordinate.  It is related to the modular parameter
\begin{equation}\label{eq:tau-SW}
  \tau = \frac{\pd t_D}{\pd t}
\end{equation}
by the identity \eqref{eq:dadz} with $c_t = 1/6$.
The genus zero free energy is defined by \eqref{eq:sgr} with
$\kappa = -1$.
The genus one free
energy can be put in the form \eqref{eq:F1} with the weight $1/2$
modular form
\begin{equation}
  \Phi = \sqrt{-\frac{\rd t}{\rd z}}\left(16(u^2-1)\right)^{1/12}.
\end{equation}
The propagator
$S$ is defined from the genus one free energy by
\begin{equation}
  \pd_uF_1 = \frac{C_u}{2}S,
\end{equation}
where the Yukawa coupling is
\begin{equation}
  C_{u} = \frac{1}{2(u^2-1)}.
\end{equation}
The propagator can also be put in the modular form
\begin{equation}\label{eq:S-SW}
  S = \frac{\kappa}{12}\left(\frac{\rd t}{\rd
      z}\right)^{-2}\wh{E}_2(\tau,\bar{\tau}) - \frac{u}{3}.
\end{equation}
which is slightly modified from  \eqref{eq:SE2}.
The remaining free energies can be solved recursively from
\eqref{eq:hae-hE2}, and are all polynomials in the propagator.
For instance, the genus two free energy is
\begin{equation}\label{eq:F2-SW}
  F_2 = C_u^2\left(\frac{5}{24}S^3 - \frac{S^2 u}{4} +
    S\left(\frac{1}{12}+\frac{u^2}{36}\right) -\frac{u}{72}-\frac{19u^3}{3240}\right).
\end{equation}
The holomorphic free energies are obtained by taking the limit
$\bar{\tau}\rightarrow -\infty$, so that
\begin{equation}
  F_1\rightarrow \CF_1 = -\log\left(-\frac{\rd t}{\rd z}\right)-\frac{1}{12}\log(16(u^2-1)),
\end{equation}
and for $g\geq 2$
\begin{equation}
  F_g(u,S) \rightarrow \CF_g(u,\CS),
\end{equation}
with the holomorphic propagator obtained by substituting $E_2$ for
$\wh{E}_2$.

We can also take the magnetic frame, where we choose $t_D$ to be the
local flat coordinate, and the modular parameter is
\begin{equation}
  \tilde{\tau} = -\frac{\pd t}{\pd t_D},
\end{equation}
which is related to \eqref{eq:tau-SW} by an S-transformation.
\begin{equation}
  \tilde{\tau} = -1/\tau.
\end{equation}
The prepotential is again defined by the special geometry relation,
while the free energies $F_{g\geq 1}$ have the same expressions such
as \eqref{eq:F2-SW} except that the propagator $S$ is obtained from
\eqref{eq:S-SW} with $t,\tau,\bar{\tau}$ replaced by
$t_D,\tilde{\tau},\bar{\tilde{\tau}}$.  

\subsection{Modular non-perturbative amplitudes and Stokes constants}

Non-perturbative amplitudes and Stokes constants of leading order
singular points of the Borel transform have been tested in
\cite{Marino:2024yme}.  Here we test the modular expression of
instanton amplitudes \eqref{eq:Fmu},\eqref{eq:SD} extensively, as well
as sub-leading singular points, mostly in the electric frame.

\begin{figure}
  \centering%
  \subfloat[No
  subtraction]{\includegraphics[height=5cm]{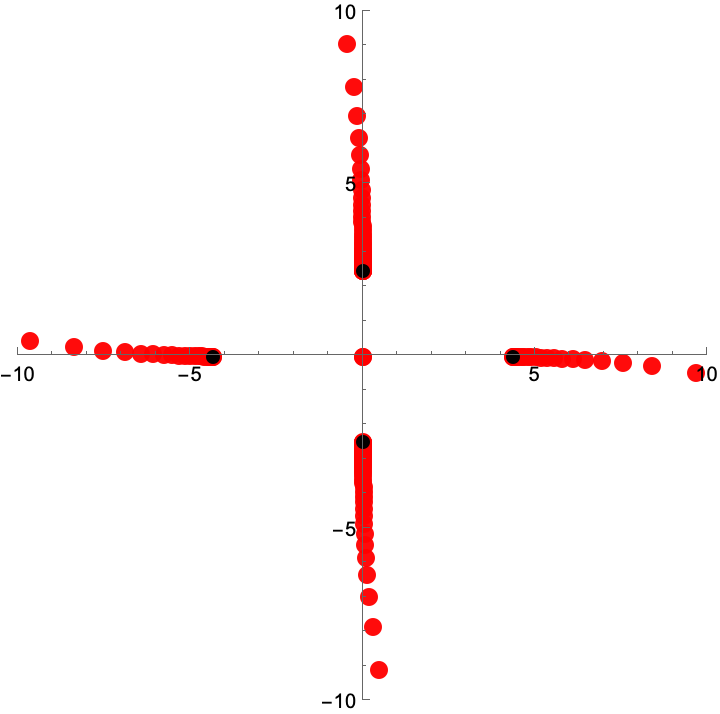}\label{fig:SWEfu1o4-brl0}}%
  \hspace{8ex}%
  \subfloat[After
  subtraction]{\includegraphics[height=5cm]{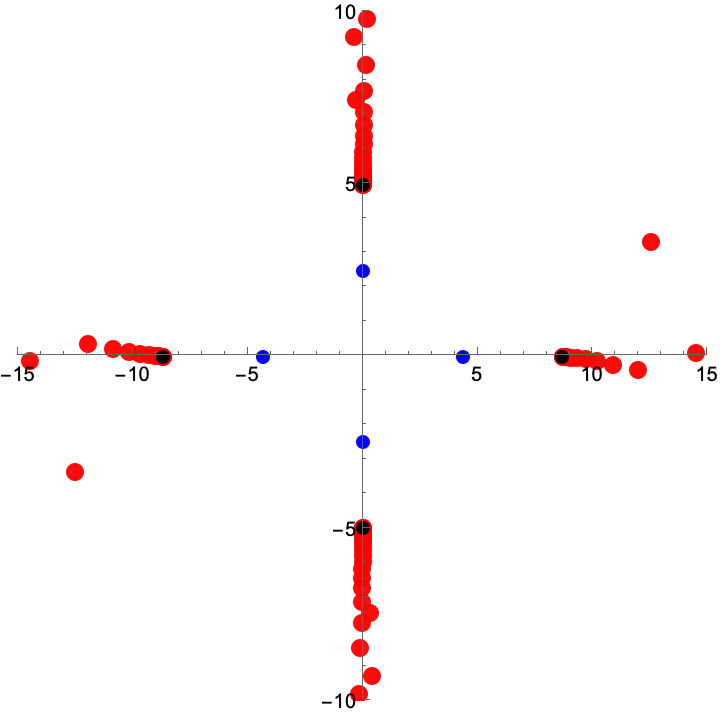}\label{fig:SWEfu1o4-brl1}}%
  \caption{Borel plane of free energies (electric frame) of
    Seiberg-Witten theory at $u=1/4$ and $\tau_2 = 200$ before (a) and
    after (b) subtracting leading order non-perturbative
    contributions.  In Fig.~(a), the black dots are the leading
    singularities $\CA_{\pm (1,0)}$ (imaginary axis),
    $\CA_{\pm (1,2)}$ (real axis).  In Fig.~(b), the black dots are
    the subleading singularities $\CA_{\pm (2,0)},\CA_{\pm (2,4)}$,
    while the blue dots mark the position of the leading singularities
    whose contributions have been successfully removed.}
  \label{fig:SWEfbrl-SC} 
\end{figure}

\begin{figure}
  \centering%
  \subfloat[$\real
  s_g^{(-1)}$]{\includegraphics[width=0.35\linewidth]{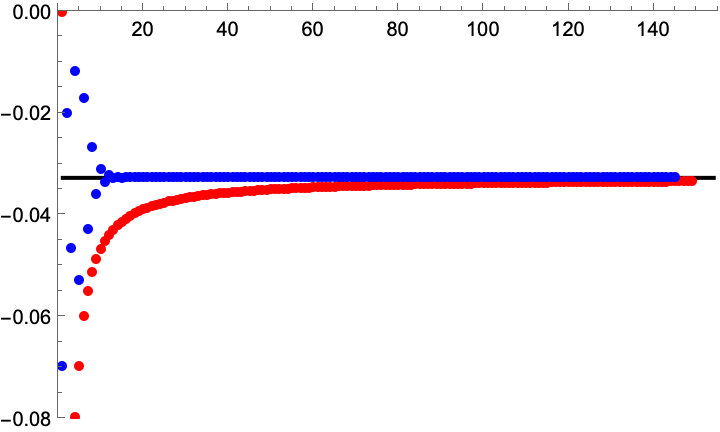}}%
  \hspace{8ex}
  \subfloat[$\imag
  s_g^{(-1)}$]{\includegraphics[width=0.35\linewidth]{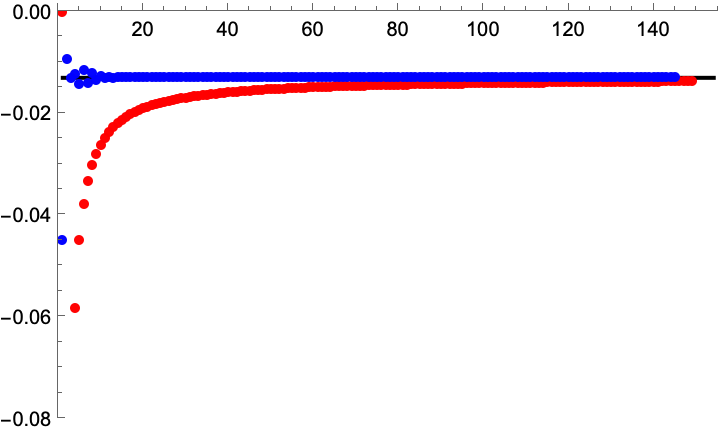}}\\%
  \subfloat[$\real
  s_g^{(0)}$]{\includegraphics[width=0.35\linewidth]{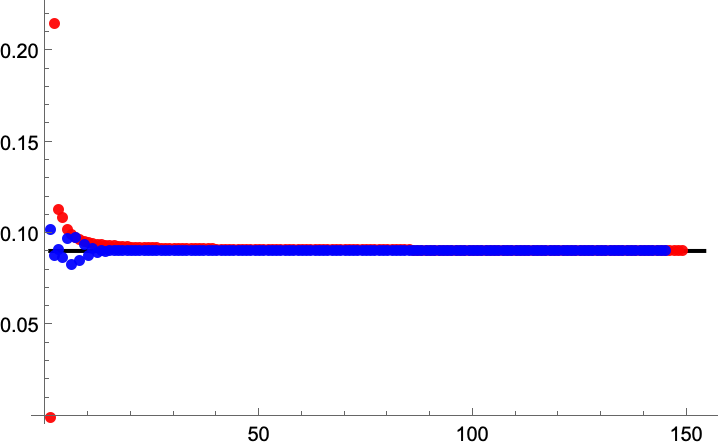}}%
  \hspace{8ex}
  \subfloat[$\imag
  s_g^{(0)}$]{\includegraphics[width=0.35\linewidth]{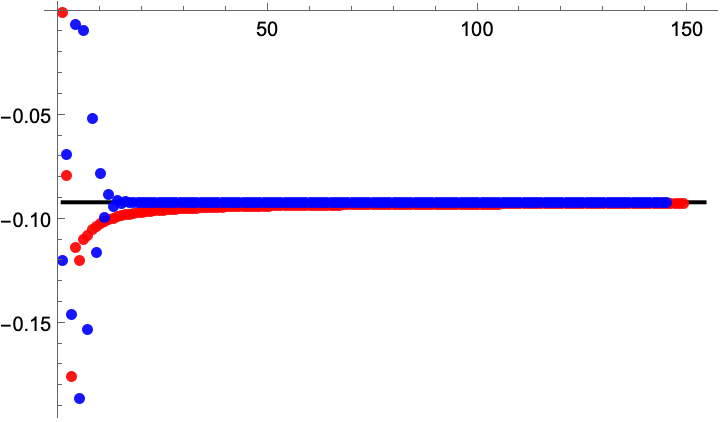}}%
  \caption{The sequences $s^{(j=-1,0)}_g$ for SW theory (electric
    frame) with $u = 1/4$ and $\tau_2 = 200$.  Red dots are numerical
    values, blue dots are results after Richardson transform, black
    line is the expected asymptotic value, the coefficient
    $a_j^{(\CA_\mu),+}$ of non-perturbative amplitude for
    $\mu=\pm (1,0)$ and $j=-1,0$, multiplied with Stokes constant
    $S_{\pm (1,0)}=1$.}
  \label{fig:SWinstcoefs-SC}
\end{figure}

In all examples, the Borel singularities are located at
\begin{equation}
  \CA_\mu = 2\pi Z_\mu
\end{equation}
where $Z_\mu$ is the central charge of certain D-brane bound state.
We first consider the strong coupling regime with small $u$.  As shown
in Fig.~\ref{fig:SWEfu1o4-brl0}, the leading order singularities are
$\CA_\mu$ with
\begin{equation}\label{eq:Amu-EfSC}
  \mu = \pm (1,0),\;\; \pm (1,2).
\end{equation}
The non-perturbative coefficients of the closest singularities can be
checked to agree with \eqref{eq:Fmu}, \eqref{eq:SD} using the
resurgence relation \eqref{eq:rel} and auxiliary sequences $s_g^{(j)}$
\eqref{eq:seq-sg}, as shown in Fig.~\ref{fig:SWinstcoefs-SC}.
Furthermore, the non-perturbative amplitudes of all four leading order
singularities can be verified by successful removal of their
contributions by subtracting \eqref{eq:Igw} from perturbative free
energies as shown in Fig.~\ref{fig:SWEfu1o4-brl1} to reveal correct sub-leading singularities located at
$\CA_{\mu'}$ with
\begin{equation}
  \mu' = \pm (2,0),\;\; \pm (2,4),
\end{equation}
which should come from $\dot{\Delta}_{\CA_\mu}^2$.
In all these tests, we
require the correct Stokes constants, which are found to be
\begin{equation}\label{eq:StConstSW-SC}
  S_{\pm (1,0)} = S_{\pm (1,2)} = 1.
\end{equation}
They agree with the DT invariants \eqref{eq:OmegaSC-SW}.

%

%

\begin{figure}
  \centering%
  \subfloat[No
  subtraction]{\includegraphics[height=5cm]{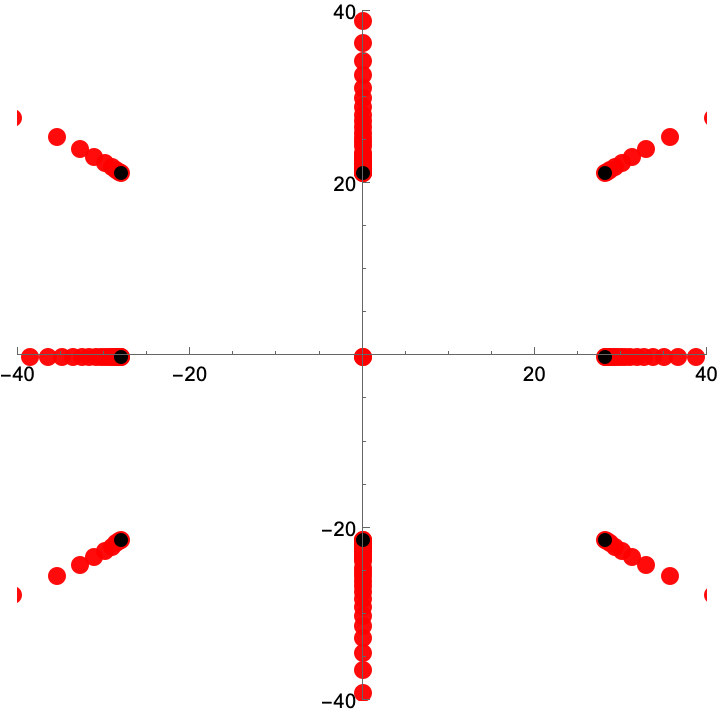}\label{fig:SWEfu10-brl0}}%
  \hspace{8ex}%
  \subfloat[After
  subtraction]{\includegraphics[height=5cm]{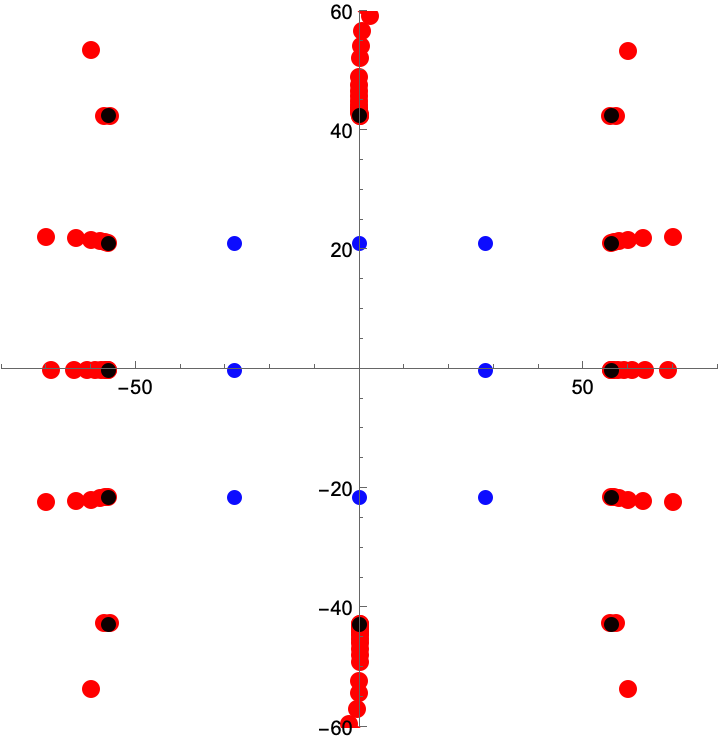}\label{fig:SWEfu10-brl1}}%
  \caption{Borel plane of free energies (electric frame) of
    Seiberg-Witten theory at $u=10$ and $\tau_2 = 300$ before (a) and
    after (b) subtracting leading order non-perturbative
    contributions.  In Fig.~(a), the black dots are the leading
    singularities $\CA_{\pm (1,0)}$ (imaginary axis),
    $\CA_{\pm (0,2)}$ (real axis), and $\CA_{\pm (1,\pm 2)}$
    (quadrants).  In Fig.~(b), the black dots are the subleading
    $\CA_{\pm (2,0)},\CA_{(\ell,\pm 4)}$ for $\ell=0,\pm 1,\pm2$,
    while the blue dots mark the position of the leading singularities
    whose contributions have been successfully removed.}
  \label{fig:SWEfbrl-WC} 
\end{figure}

\begin{figure}
  \centering%
  \subfloat[$\real s_g^{(-1)}$]{\includegraphics[width=0.35\linewidth]{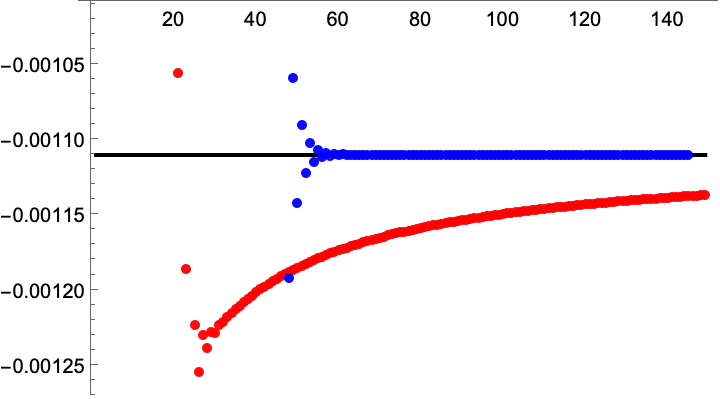}}%
  \hspace{8ex}
  \subfloat[$\imag s_g^{(0)}$]{\includegraphics[width=0.35\linewidth]{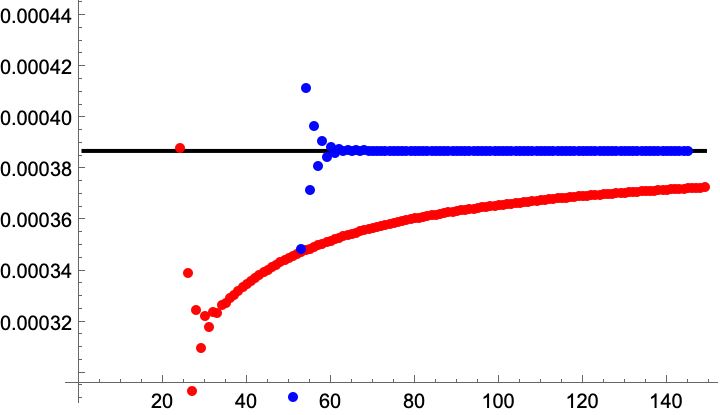}}%
  \caption{The sequences $s^{(j=-1,0)}_g$ for SW theory (electric
    frame) with $u = 10$ and $\tau_2 = 300$.  Red dots are numerical
    values, blue dots are results after Richardson transform, black
    line is the expected asymptotic value, the coefficient
    $a_j^{(\CA_\mu),+}$ of non-perturbative amplitude for
    $\mu=\pm (1,0)$ and $j=-1,0$, multiplied with Stokes constant
    $S_{\pm (1,0)}=1$. Note that $s^{(-1)}_g$ are purely real, while
    $s^{(0)}_g$ are purely imaginary.}
  \label{fig:SWinstcoefs-WC}
\end{figure}

Next, consider the weak coupling regime with large $u$.
As shown in Fig.~\ref{fig:SWEfu10-brl0}, the leading order
singularities are $\CA_\mu$ with
\begin{equation}\label{eq:Amu-EfWC}
  \mu = \pm (1, 0),\;\; \pm (0,2), \;\; \pm (1,\pm 2).
\end{equation}
The non-perturbative coefficients of the closest singularities can
again be checked to agree with \eqref{eq:Fmu}, \eqref{eq:SD} using the
resurgence relation \eqref{eq:rel} and auxiliary sequences $s_g^{(j)}$
\eqref{eq:seq-sg}, as shown in Fig.~\ref{fig:SWinstcoefs-WC}, while
non-perturbative amplitudes of all six leading order singularities can
be verified by successful removal of their contributions as shown in
Fig.~\ref{fig:SWEfu10-brl1} to reveal correct sub-leading
singularities in the Borel plane, located at $\CA_{\mu'}$ with
\begin{equation}
  \mu' = \pm (2,0),\;\; (\ell,\pm 4),\;\text{for}\;\ell=0,\pm 1,\pm 2,
\end{equation}
which should come from $\dot{\Delta}_{\CA_\mu}^2$ and
$\dot{\Delta}_{\CA_{\pm (1,\pm 4)}}$.  In all these tests, we require the
correct Stokes constants, which are found to be
\begin{equation}\label{eq:StConstSW-WC}
  S_{\pm (1,0)} = S_{\pm (1,\pm 2)} = 1,\quad
  S_{\pm (0,2)} = -2.
\end{equation}
They agree with the DT invariants \eqref{eq:OmegaWC-SW}.

%

We now argue that using the property of monodromy invariance of Stokes
constants, we can generate a large subset if not all of Stokes
constants which reproduce the entire BPS spectrum.  In the strong
coupling regime, there are no local monodromy action.  There are no
cusp points inside the strong coupling regime, and to circle around
the cusp points $\pm 1$ on the boundary, one will need to cross the
wall of marginal stability, and thus the local monodromy argument
cannot be applied. In any case, \eqref{eq:StConstSW-SC} already
reproduce the full BPS spectrum in the strong coupling regime.

%

\begin{figure}
  \centering%
  \includegraphics[height=5cm]{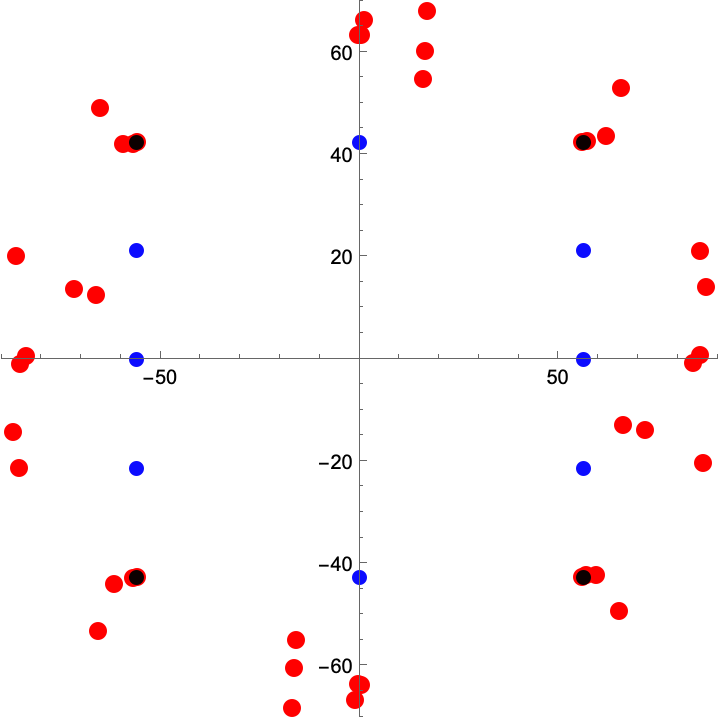}\label{fig:SWEfu10-brl2p}%
  \caption{Borel plane of free energies (electric frame) of
    Seiberg-Witten theory at $u=10$ and $\tau_2 = 300$ after removing
    some of the sub-leading order non-perturbative contributions.  Due
    to numerical precision, we cannot remove $\CA_{\pm (2,\pm 4)}$,
    which are marked by black dots, while
    $\CA_{\pm (2,0)}, \CA_{\pm (0,4)}, \CA_{\pm (1,\pm 4)}$ are all
    successfully removed, which are marked by blue dots.}
  \label{fig:SWEfbrl-WC2p} 
\end{figure}

In contrast, in the weak coupling regime, there is the cusp point at
$u=\infty$ inside the stability chamber, as also indicated in
Fig.~\ref{fig:SW-wall}, and the associated monodromy matrix is
$M_\infty$ in \eqref{eq:M-SW}, which generates a subgroup $\IZ$ of the
monodromy group $\Gamma(2)$.  The two singular points
$\CA_{\pm (0,2)}$ form a closed finite orbit of the subgroup.  On the
other hand, the singular points $\CA_{\pm (1,0)}, \CA_{\pm (1,2)}$
generate an infinite orbit consisting of $\CA_{(\pm 1,2\ell)}$ for
$\ell\in \IZ$ and they all share the same Stokes constant
\begin{equation}\label{eq:S2l-SW}
  S_{(\pm 1,2\ell)} = 1.
\end{equation}
We thus have also reproduced the full BPS spectrum
\eqref{eq:OmegaWC-SW} in the weak coupling regime.  In fact,
\eqref{eq:S2l-SW} can be further checked by successful removal of
contributions from some of the sub-leading singularities in the Borel
plane, as shown in Fig.~\ref{fig:SWEfbrl-WC2p}, including the
contributions from $\CA_{\pm (2,0)}, \CA_{\pm (0,4)}$ as well as
$\CA_{\pm (1,\pm 4)}$, the latter of which require the correct Stokes
constants $S_{\pm (1,\pm 4)}=1$. Note that by successfully removing
contributions from $\CA_{\pm (2,0)}, \CA_{\pm (0,4)}$, we have also
verified the non-perturbative amplitudes beyond the 1-instanton level
\eqref{eq:DelDnZ} as well as the formula \eqref{eq:Sbar} for Stokes
constants, away from the holomorphic limit with the modular derivative
\eqref{eq:SD}.

\begin{figure}
  \centering%
  \includegraphics[height=5cm]{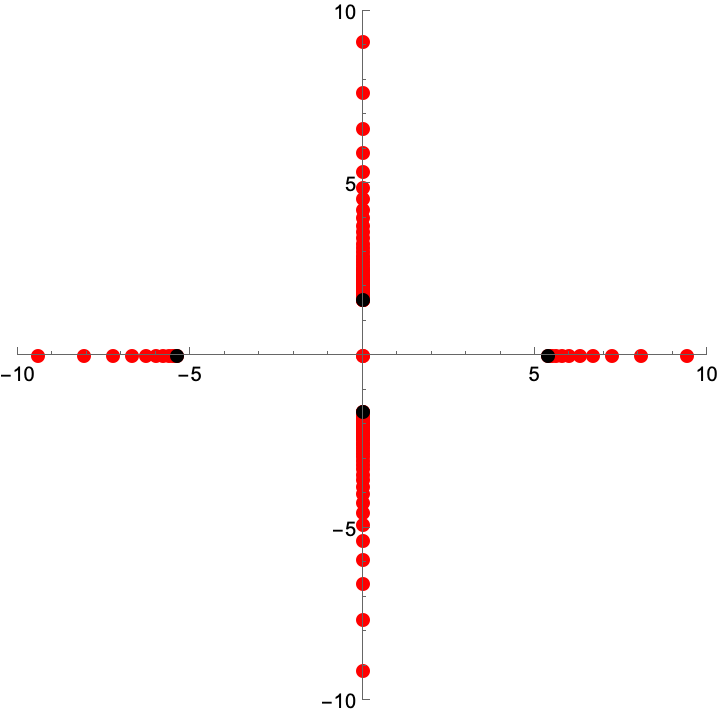}\label{fig:SWMfu1o2-brl}
  \caption{Borel plane of free energies (magnetic frame) of
    Seiberg-Witten theory at $u=1/2$ and $\tau_2 = 300$. The black
    dots are the leading singularities $\CA_{\pm (1,0)}$ (imaginary
    axis), $\CA_{\pm (1,2)}$ (real axis).}
  \label{fig:SWMfbrl-SC} 
\end{figure}

\begin{figure}
  \centering%
  \subfloat[$\real s_g^{(-1)}$]{\includegraphics[width=0.35\linewidth]{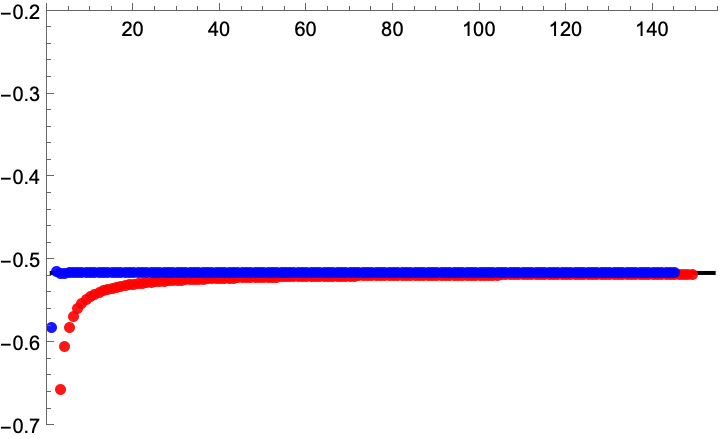}}%
  \hspace{6ex}
  \subfloat[$\imag s_g^{(0)}$]{\includegraphics[width=0.35\linewidth]{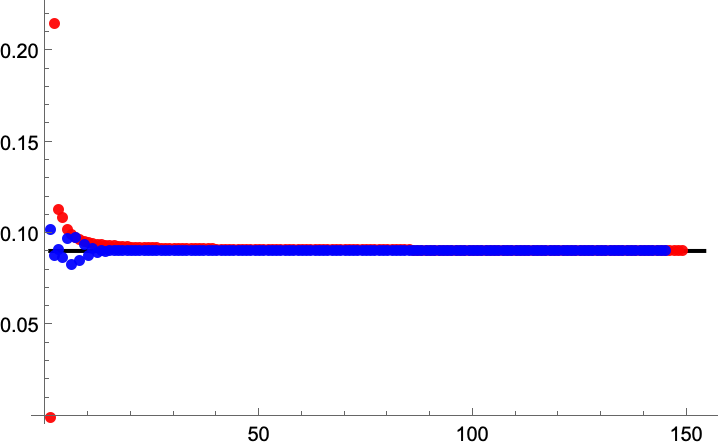}}%
  \caption{The sequences $s^{(j=-1,0)}_g$ for SW theory (magnetic
    frame) with $u = 1/2$ and $\tau_2 = 300$.  Red dots are numerical
    values, blue dots are results after Richardson transform, black
    line is the expected asymptotic value, the coefficient
    $a_j^{(\CA_\mu),+}$ of non-perturbative amplitude for
    $\mu=\pm (1,0)$ and $j=-1,0$, multiplied with Stokes constant
    $S_{\pm (1,0)}=1$. Note that $s^{(-1)}_g$ are purely real, while
    $s^{(0)}_g$ are purely imaginary.}
  \label{fig:SWinstcoefs-MfSC}
\end{figure}

With the modular formalism of the non-perturbative amplitude, the
choice of the frame no longer makes a big difference. Nevertheless, we
have also checked the non-perturbative amplitudes in other frames.  To
formulate the free energies in a different frame, we simply perform a
modular transformation.  For instance, in the magnetic frame, where
$t_D$ serves as the local flat coordinate, the free energie $F_g$ are
obtained by an $S$-transformation
\begin{equation}
  \tau \rightarrow \tau_D = -\frac{1}{\tau}.
\end{equation}
We perform similar tests of the modular expressions of
non-perturbative amplitudes in the magnetic frame, in the strong
coupling regime.  The Borel plane is given in
Fig.~\ref{fig:SWMfbrl-SC}, where the leading order singular points are
still $\CA_\mu$ with $\mu$ given in \eqref{eq:Amu-EfSC}, and the
non-perturbative coefficients of the closest singularity can be
verified by the resurgence relation as shown in
Fig.~\ref{fig:SWinstcoefs-MfSC} with the Stokes constants
\eqref{eq:StConstSW-SC}.

\section{Example: 4d $SU(2)$ $N_f=4$ theory}
\label{sec:Nf4}

\subsection{Perturbative free energies}

The next example we consider is the 4d $SU(2)$ massless SQCD with
$N_f=4$ hypermultiplets. This theory has a single Coulomb modulus $t$
and it is known to be conformal in the massless limit
\cite{Seiberg:1994aj,Seiberg:1994rs}, so that the effective gauge
coupling $\tau$ does not run and the relation \eqref{eq:tau} is
replaced by the simple formula
\begin{equation}
  \tau = \frac{t_D}{t}.
\end{equation}
When coupled to
the Omega background in the self-dual limit, the gauge theory has a
topological string interpretation, which is formulated on the mirror
curve \cite{Huang:2011qx}
\begin{equation}
  y^2 = x(x-u)(x-uq_{0}),
\end{equation}
where the modulus $u$ is related to the Coulomb modulus $t$ by
\cite{Seiberg:1994aj}
\begin{equation}
  t = \sqrt{u/2},
\end{equation}
and $q_{0} = \exp(2\pi\ri \tau_{0})$ is the gauge coupling in the UV.
The mirror curve is elliptic and it can be put in the Weierstrass form
with \cite{Huang:2011qx}  
\begin{equation}
  \begin{aligned}
    g_2 = &\frac{1}{12}(q_0^2-q_0+1)u^2,\\
    g_3 = &\frac{1}{432}(2q_0^3-3q_0^2-3q_0+2)u^3.
  \end{aligned}
\end{equation}
The relation between the gauge coupling $\tau$ in the IR and the
coupling $q_0$ in the UV can then be established through the
$J$-invariant \eqref{eq:Jinv}.  Alternatively, the superconformal
theory can be obtained in certain field theory limit of topological
string on Enriques Calabi-Yau threefold \cite{Grimm:2007tm}.

The preportential of the theory is defined through the geometry
relation \eqref{eq:sgr} with $\kappa = 1$, and it is simply
\cite{Dorey:1996bn} 
\begin{equation}
  \CF_0 = -\pi\ri\kappa \tau t^2.
\end{equation}
The genus one free energy is given by \cite{Huang:2011qx}
\begin{equation}\label{eq:F1-Nf4}
  F_1 = -\frac{1}{2}\log(\tau_2|t|^2),
\end{equation}
which is consistent with \eqref{eq:F1} where the weight $1/2$ modular
form is $\sqrt{t}$.  Higher genera free energies can be calculated
through the holomorphic anomaly equations.  This can be done in two
different ways. One can either turn on the mass deformation with for
instance $m_1 = m_2 = m$, $m_3=m_4=0$ and solve the holomorphic
anomaly equations \eqref{eq:hae-hE2}, where the holomorphic
ambiguities can be fixed by the usual gap condition
\cite{Huang:2011qx}.  Alternatively, one can stay in the massless
limit.  The free energies have well-defined scaling behavior and can
be written as \footnote{The functions $H$ here are $f$ in
  \cite{Huang:2011qx}.}
\begin{equation}\label{eq:FgHg}
  F_g = (g-1)^{-1}\at^{2-2g} H_g(\tau,\bar{\tau}),\quad g\geq 2.
\end{equation}
The holomorphic anomaly equations \eqref{eq:hae-hE2} then reduce to
the simplified equations for the $t$-independent functions $H_g$
\cite{Huang:2011qx}
\begin{equation}\label{eq:hae-H}
  \frac{\pd H_g}{\pd \wh{E}_2} = \frac{\kappa(g-1)}{6}
  \left(\left(g-\frac{3}{2}\right)H_{g-1} + \sum_{g'=1}^{g-1}H_{g'} H_{g-g'}\right).
\end{equation}
%
It is then argued in \cite{Huang:2011qx} that $H_g$ for $g\geq 2$ is
an element of the ring of almost holomorphic modular forms generated
by $\wh{E}_2,E_4,E_6$ of modular weight $2g-2$. Therefore, we can make
the ansatz for $H_g$
\begin{equation}
  H_g = \sum_{k=0}^{g-1} P_{g,k}(E_4,E_6)\wh{E}_2^k,
\end{equation}
where $P_{g,0}$ is the holomorphic ambiguity which is a polynomial in
$E_4$ and $E_6$ only,
\begin{equation}
  P_{g,0} = \sum_{2k_4+3k_6=g-1}
  c_{k_4,k_6}E_4^{k_4}E_6^{k_6}.
\end{equation}
By plugging the ansatz into the holomorphic anomaly equations
\eqref{eq:hae-H}, the latter can be solved recursively with respect to
the genus $g$, as long as the holomorphic ambiguity $P_{g,0}$ can be
fixed in a suitable way.

In usual models of topological string theory, holomorphic ambiguities
are fixed by exploiting the gap conditions of the free energies.
Although the 4d massless $N_f=4$ theory also has gap conditions, as we
will see in Section~\ref{sec:gap}, they are not very useful for fixing
the holomorphic ambiguity $H_{g,0}$.  Instead, inspired by
\cite[Sec.~6]{Grimm:2007tm}, we propose \footnote{We thank Marcos
  Marino for suggesting this idea in the early stage of the project.}
to fix the holomorphic ambiguities by comparing the holomorphic limit
of $F_g$, which is obtained through simply relacing $\wh{E}_2$ by
$E_2$, with the genus expansion of the Nekrasov instanton partition
function \cite{Nekrasov:2002qd}.
%
%
In this way, we have calculated free energies up to genus $222$, using
the Nekrasov instanton partition function up to degree $37$.  We
tabulate some of the results of $F_g$ in the following:
\begin{equation}
  \begin{aligned}
    &F_2 = \frac{\wh{E}_2}{2^5 a^2},\\
    &F_3 = \frac{1}{2^7 a^4}\left(\frac{2}{3}\wh{E}_2^2 +\frac{1}{3}E_4\right),\\
    &F_4 = \frac{1}{2^9a^6}\left(\frac{11}{12}\wh{E}_2^3+\frac{4}{3}\wh{E}_2E_4+\frac{7}{12}E_6\right),\\
    &\ldots
  \end{aligned} 
\end{equation}
whose holomorphic limit agree with \cite{Grimm:2007tm,Huang:2011qx}.


The free energies $F_g$ thus calculated are in the electric frame. We
can switch to a different frame by the modular transformation
\begin{equation}
    \begin{pmatrix}
        a_D \\ a
    \end{pmatrix}  \rightarrow 
    \begin{pmatrix}
        \tilde{a}_D \\ \tilde{a}
    \end{pmatrix} = \gamma \cdot 
    \begin{pmatrix}
        a_D \\ a
    \end{pmatrix},\quad 
    \gamma = \begin{pmatrix}
        a & b \\ c & d
    \end{pmatrix}\in SL(2,\IZ),
\end{equation}
which indicates the familiar modular transformation on the modular parameter
\begin{equation}
    \tau \rightarrow \tilde{\tau} = \gamma\cdot\tau = \frac{a\tau+b}{c\tau + d}.
\end{equation}
In particular, we can switch to the magnetic frame by
\begin{equation}
    \gamma = \begin{pmatrix}
        0 & -1\\1 & 0
    \end{pmatrix},\quad \tau\rightarrow \tilde{{\tau}} = -1/\tau.
\end{equation}
The free energies are
\begin{equation}
  \begin{aligned}
    &F^{\rm m}_2 = \frac{1}{2^5 a_D^2}\wh{E}_2(\tilde{\tau},\bar{\tilde{\tau}}),\\
    &F^{\rm m}_3 = \frac{1}{2^7 a_D^4}\left(\frac{2}{3}\wh{E}_2^2(\tilde{\tau},\bar{\tilde{\tau}}) +\frac{1}{3}E_4(\tilde{\tau})\right),\\
    &F^{\rm m}_4 =
    \frac{1}{2^9a_D^6}\left(\frac{11}{12}\wh{E}_2^3(\tilde{\tau},\bar{\tilde{\tau}})
      +\frac{4}{3}\wh{E}_2(\tilde{\tau},\bar{\tilde{\tau}})E_4(\tilde{\tau})+\frac{7}{12}E_6(\tilde{\tau})\right),\\
    &\ldots
  \end{aligned}
\end{equation}

\subsection{BPS spectrum}

The 4d $N=2$ $SU(2)$ massless $N_f=4$ theory has flavor symmetry group
$SO(8)$, and a one dimensional Coulomb branch moduli space
parametrised by $u$. 
The Coulomb branch has one
singular point located at \cite{Seiberg:1994aj}
\begin{equation}
  u=0,
\end{equation}
out of which comes a branch cut, but
there is \emph{no} wall of marginal stability.  The condition of the
stability wall
\begin{equation}
  \imag a_D/ a = 0
\end{equation}
is equivalent to the boundary of the upper half plane,
\begin{equation}
  \imag \tau>0.
\end{equation}
Thus, over the entire Coulomb branch, the BPS spectrum is the same,
and it consists of the following \cite{Ferrari:1997gu}: 
\begin{itemize}
\item Hypermultiplets of charges $(p,q)$ where $p,q$ are coprime
  integers\footnote{For instance, there are hypermultiplets of charges
    $(1,0)$, $(1,1)$, $(0,1)$, but not of charges $(2,0)$, $(2,2)$,
    $(0,2)$.}.  They include fundamental quarks of charge $\pm (0,1)$
  which transform as $\md{8}_v$ of the flavor symmetry group $SO(8)$,
  and monoples of charge $\pm (1,0)$ which transform as either
  $\md{8}_s$ or $\md{8}_c$ of $SO(8)$.  Each hypermultiplet, after
  factoring out the universal hypermultiplet, is counted as a BPS
  state of spin 0 and contributes one to the DT invariant.  The total
  DT invariant for each charge $(p,q)$ is
  \begin{equation}\label{eq:BPS-hyp}
    \Omega(p,q) = 8.
  \end{equation}
\item Vectormultiplets of charges $(2p,2q)$ where $p,q$ are coprime
  integers.  They include the W-bosons of charge $\pm (0,2)$, which
  transform as singlet of the flavor symmetry group.  Each
  vectormultiplet after factoring out the universal hypermultiplet is
  counted as a BPS state with spin 1/2 and contributes $-2$ to the DT
  invariant.  Therefore, for each charge $(2p,2q)$, the DT invariant
  is 
  \begin{equation}\label{eq:BPS-vec}
    \Omega(2p,2q) = -2.
  \end{equation}
\end{itemize}

The full BPS spectrum is in fact manifestly self-dual, invariant under
$SL(2,\IZ)$ transformations \cite{Ferrari:1997gu}.
Conversely, assuming $SL(2,\IZ)$ invariance, the entire BPS spectrum
can be generated by two states, the hypermultiplet of charge
$\pm (1,0)$ and the vectormultiplet of charge $\pm
(0,2)$. 


\subsection{Gap conditions and distinguished frame non-perturbative
  free energy}
\label{sec:gap}

To find possible non-perturbative corrections to \eqref{eq:F-pert}, we
first consider a simple scenario, the holomorphic limit in the
electric frame, where the free energies $F_g$ reduce to $\CF_g$ which
are polynomials of $E_2,E_4,E_6$.  In addition, we also take the limit
\begin{equation}
  \tau \rightarrow \ri\infty.
\end{equation}
The Eisenstein series are
\begin{equation}
  E_k(\tau) \sim 1 + \CO(\re^{2\pi\ri\tau}),
\end{equation}
and we find gap-like behavior
\begin{equation}\label{eq:gap-e}
  \CF_g(t,\tau) \sim
  \frac{(-1)^{g-1}8(1-2^{-2g})B_{2g}}{2g(2g-2)t^{2g-2}} +
  \CO(\re^{2\pi\ri t_D/t}),\quad g\geq 2.
\end{equation}
In fact, this originates from nothing else but the perturbative free
energy from gauge theory, or the genuine gap condition of topological
string on Enriques Calabi-Yau threefold \cite{Grimm:2007tm}.
Similarly, if we take the limit
\begin{equation}
  \tilde{\tau}\rightarrow \ri 0^+
\end{equation}
in the magnetic frame, we find
\begin{equation}\label{eq:gap-m}
  \CF_g^{\rm m}(t_D,\tau)  \sim
  \frac{(-1)^{g-1}8(1-2^{-2g})B_{2g}}{2g(2g-2)t_D^{2g-2}} +
  \CO(\re^{2\pi\ri t/t_D}),\quad g\geq 2.
\end{equation}
It is clear from these behaviors that $\CF_g\sim (2g)!$ for large $g$.

The gap conditions are indicative of the leading order
non-perturbative corrections.  Without loss of generality, let us work
in the electric frame and define
\begin{equation}
  \CF^{\rm gap}(t;g_s) = \sum_{g=2}^\infty \CF_g^{\rm gap}(t) g_s^{2g-2}
\end{equation}
where we have collected the leading order contributions in
\eqref{eq:gap-e} with
\begin{equation}
  \CF_g^{\rm gap}(t)  = \frac{(-1)^{g-1}8(1-2^{-2g})B_{2g}}{2g(2g-2)t^{2g-2}}.
\end{equation}
Using the identity for the Bernoulli numbers,
\begin{equation}
  B_{2g} = (-1)^{g+1}\frac{2(2g)!}{(2\pi)^{2g}}\sum_{k=1}^\infty\frac{1}{k^{2g}},
\end{equation}
the leading contribution free energy $\CF^{\text{gap}}$ can be written
as
\begin{align}
  \CF^{\text{gap}}(t;g_s) = \frac{16t^2}{g_s^2}\sum_{g=2}^\infty\sum_{k=1}^\infty
  &\left(
    (2g-2)!\Big(\frac{g_s}{2\pi k t}\Big)^{2g}+(2g-3)!\Big(\frac{g_s}{2\pi k t}\Big)^{2g}\right.
    \nn
  &\phantom{=}\left.-(2g-2)!\Big(\frac{g_s}{4\pi k t}\Big)^{2g}-(2g-3)!\Big(\frac{g_s}{4\pi k t}\Big)^{2g}\right).
\end{align}
Exchange the order of the two infinite sums, and evaluate the sum over
$g$ using the formulae
\begin{subequations}
\begin{align}
  &\sum_{g=2}^\infty (2g-2)!z^{2g} =
  z^4\int_0^\infty\rd\zeta\,\re^{-\zeta}\frac{\zeta^2}{1-(z\zeta)^2},\\
  &\sum_{g=2}^\infty (2g-3)!z^{2g} =
  z^4\int_0^\infty\rd\zeta\,\re^{-\zeta}\frac{\zeta}{1-(z\zeta)^2},
\end{align}
\end{subequations}
we find that $\CF^{\text{gap}}(a;g_s)$ has the following integral
representation
\begin{equation}\label{eq:bsum}
  \CF^{\text{gap}}(t;g_s) \sim 16t^2g_s\sum_{k=1}^\infty(1/2\pi k t)^4
  \int_0^{\re^{\ri\arg g_s}\infty}\rd\zeta\,\re^{-\zeta/g_s}
  \left(\frac{\zeta/g_s+(\zeta/g_s)^2}{1-(\zeta/2\pi k t)^2} - \frac{1}{16}
    \frac{\zeta/g_s+(\zeta/g_s)^2}{1-(\zeta/4\pi k t)^2}
  \right).
\end{equation}
Comparing with \eqref{eq:Bsum}, this integral representation can be
interpreted as Borel resummation
\begin{equation}
  \mr{S}^{(\theta)}\CF^{\text{gap}}(t;g_s) = g_s^{-1}\int_0^{\re^{\ri\theta}\infty} \rd \zeta \,
  \re^{-\zeta/g_s} \mr{B}\CF^{\text{gap}}(t;\zeta),
\end{equation}
where the Borel transform is
\begin{equation}\label{eq:BF-gap}
  \mr{B}\CF^{\text{gap}}(t;\zeta) = 16t^2g_s^2\sum_{k=1}^\infty(1/2\pi k t)^4
  \left(\frac{\zeta/g_s+(\zeta/g_s)^2}{1-(\zeta/2\pi k t)^2} - \frac{1}{16}
    \frac{\zeta/g_s+(\zeta/g_s)^2}{1-(\zeta/4\pi k t)^2}
  \right).
\end{equation}

The Borel transform $\mr{B}\CF^{\text{gap}}(\zeta)$ \eqref{eq:BF-gap}
has a series of singular points along the Stokes ray with inclination
angle $\theta = \arg a$, located at $\zeta = k \CA$ for $k=1,2,\ldots$
with
\begin{equation}\label{eq:Aa}
  \CA= 2\pi t.
\end{equation}
The Stokes discontinuity across the associated Stokes ray can be
calculated to be
%
\begin{align}
  \text{Disc}_{\theta}\CF^{\text{gap}}(t;g_s) =
  &32\pi\ri t^2g_s \sum_{k=1}^\infty
    (1/2\pi k t)^4\times\nn
  &\left\{\Res_{\zeta = 2\pi k t}
    \frac{(\zeta/g_s+(\zeta/g_s)^2)\re^{-\zeta/g_s}}{1-(\zeta /2\pi k t)^2}
    - \frac{1}{16} \Res_{\zeta = 4\pi k t}
    \frac{(\zeta/g_s+(\zeta/g_s)^2)\re^{-\zeta/g_s}}{1-(\zeta /4\pi k t)^2}
    \right\}\nn=
  &\frac{8}{2\pi\ri}\left\{
    \sum_{k=1}^\infty \frac{1}{(2k-1)^2}\left(\CA_{2k-1}/g_s+1\right)\re^{-\CA_{2k-1}/g_s}
    \right\}.
    \label{eq:DiscWF0}
\end{align}
where we have defined
\begin{equation}
  A_\ell = \ell A.
\end{equation}
It indicates that we have contributions from infinitely many
non-perturbative sectors, and in each sector, the transseries is
truncated.

By comparing with
\eqref{eq:Disc-sum},\eqref{eq:Disc-Del},\eqref{eq:alien}, and keeping
in mind that alien derivatives annihilate truncated power series, we
find that eq.~\eqref{eq:DiscWF0} indicates that
\begin{equation}\label{eq:Fgap-np}
  \dot{\Delta}_{\CA_\ell}\CF^{\text{gap}} = \frac{8}{2\pi\ri \ell^2}
  (1+\CA_\ell/g_s)\re^{-\CA_\ell/g_s}, \quad \text{for odd }\ell,
\end{equation}
and vanishes for even $\ell$.
This agrees with \eqref{eq:Fmu},\eqref{eq:Sbar},\eqref{eq:DelF-A}, as
long as the Stokes constants are chosen to be
\begin{equation}\label{eq:S1S2}
  S_{\CA_1} = 8, \quad S_{\CA_2} = -2.
\end{equation}
The Borel singularities $\CA_1, \CA_2$ carry charges $(0,1)$ and
$(0,2)$ respectively, and the associated Stokes constants
\eqref{eq:S1S2} agree precisely with the BPS multiplicites
\eqref{eq:BPS-hyp} and \eqref{eq:BPS-vec} for the fundamental quarks
and the W-bosons.

\subsection{Generic non-perturbative free energy}

Eq.~\eqref{eq:Fgap-np} is only the leading order non-perturbative
contributions to the free energies.  To find additional contributions,
we evaluate the full free energies $F_g$ and apply the resurgence
analysis.  We only work in the electric frame, and set $t=1$ as it is
only a scaling factor in the free energies.

\begin{figure}
  \centering%
  \subfloat[No subtraction]%
  {\includegraphics[height=5.5cm]{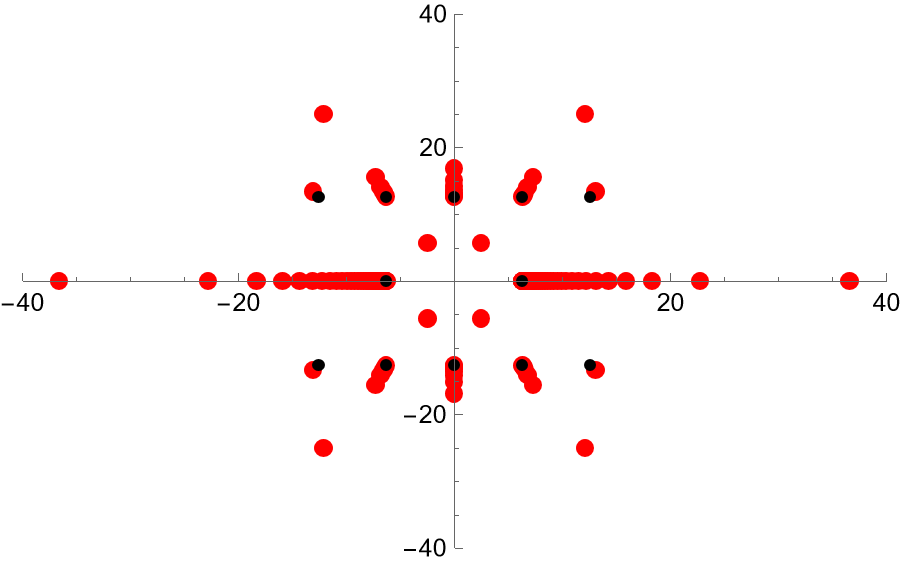}\label{fig:Nf4tau2i_nonremove}}%
  \hspace{8ex}%
  \subfloat[After subtraction]
  {\includegraphics[height=5cm]{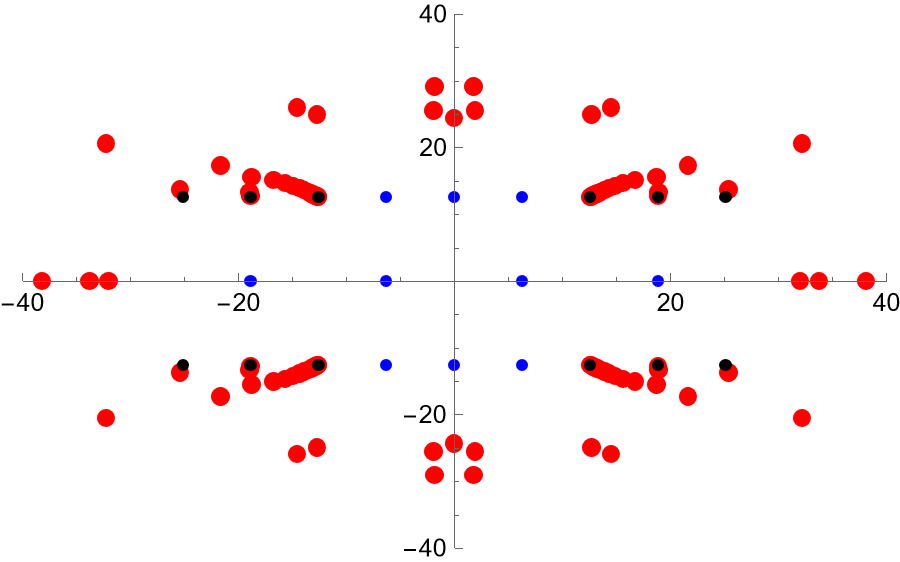}\label{fig:Nf4tau2i_remove}}%
  \caption{Borel plane of holomorphic free energies (electric frame)
    of 4d massless $N_f=4$ theory at $\tau = 2\ri$ with $t=1$ before (a)
    and after (b) removing non-perturbative contributions
    $\CA_{\pm (1,0)}$, $\CA_{\pm (3,0)}$, $\CA_{\pm (0,1)}$ and
    $\CA_{\pm (1,\pm 1)}$.  In Fig.~(a), the black dots are the leading
    singularities $\CA_{\pm (1,0)}$ (imaginary axis),
    $\CA_{\pm (0,1)}$ (real axis), and $\CA_{\pm (1,\pm 1)}$,
    $\CA_{\pm (1,\pm 2)}$ (quadrants).  In Fig.~(b), the black dots
    are the subleading $\CA_{\pm (1,\pm \ell)}$ for
    $\ell=\pm 2,\pm 3,\pm4$, while the blue dots mark the position of
    the singularities whose contributions have been successfully
    removed.}
  \label{fig:2-WC} 
\end{figure}

\begin{figure}
  \centering
  \includegraphics[height=4.5cm]{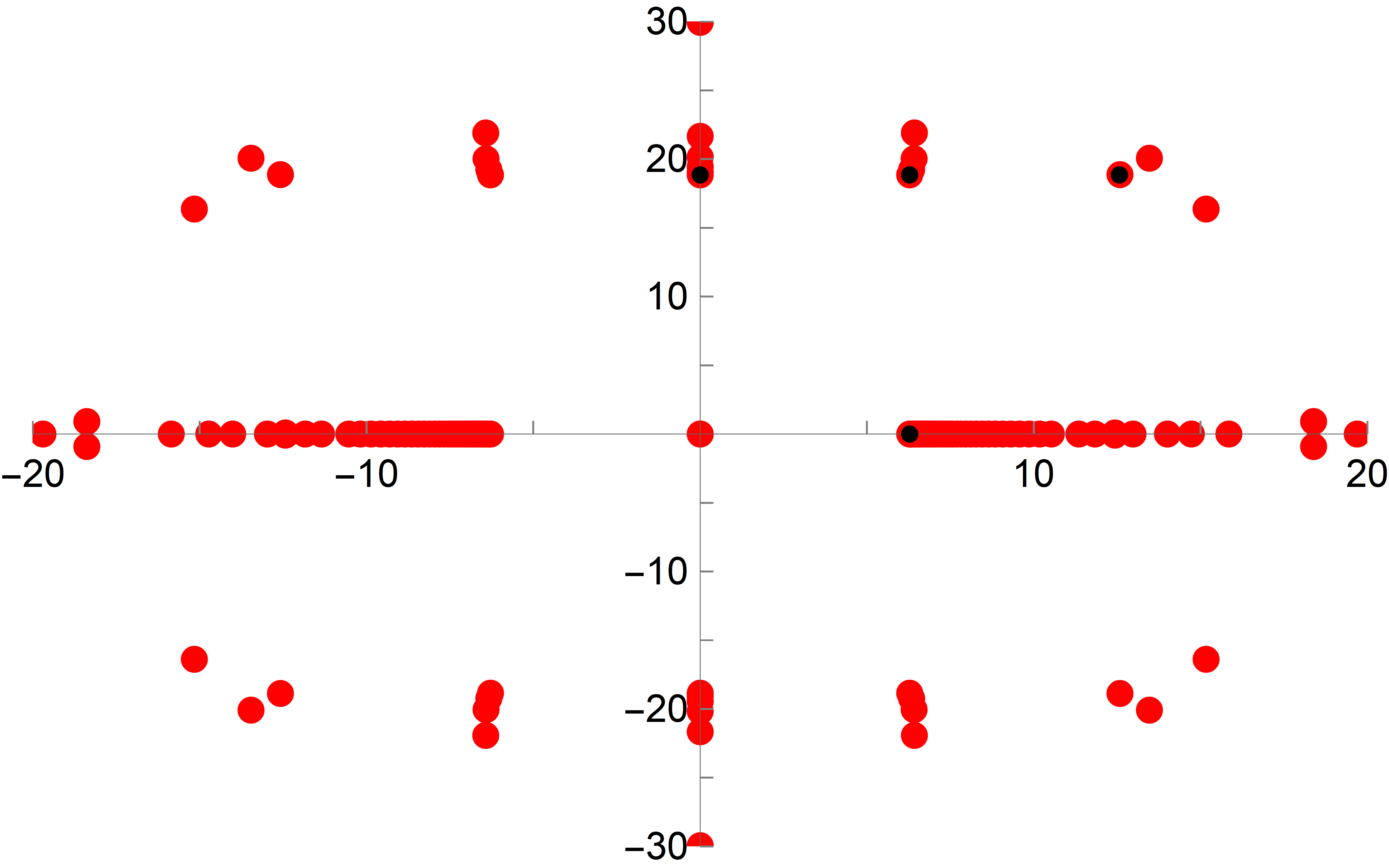}
  \caption{Borel plane of non-holomorphic perturbative free energy
    (electric frame) of 4d massless $N_f=4$ theory at $\tau = 3\ri$
    and $\tau_2 = 3$ with $t=1$.  The black dots are the leading
    singularities $\CA_{(0,1)}$, $\CA_{(1,0)}$, $\CA_{(1,1)}$, and
    $\CA_{1,2}$.}
  \label{fig:brl-nonholtau3i} 
\end{figure}

We first consider the holomorphic limit, where we find that the
singularities of the Borel transform are located at
\begin{equation}\label{eq:A-gamma-Nf4}
  \CA_{\gamma} = 2\pi (p t_D + q t).
\end{equation}
For instance, as shown in Fig.~\ref{fig:Nf4tau2i_nonremove}, the
visible singularities are of the form \eqref{eq:A-gamma-Nf4} with
\begin{equation}
  \gamma = \pm (0,1),\;\pm (1,\ell),\;\ell\in \IZ.
\end{equation}
The non-perturbative amplitudes and the Stokes constants can be
checked by successful removal of the non-perturbative contributions
from the free energy $\CF_g$ and the disappearance of the singular
points in the Borel plane.  The contributions from $\CA_{\pm (0,1)}$
are simple as the electric frame is the distinguished frame so that the
non-perturbative ampitudes are of the truncated form, which is simply
\eqref{eq:Fgap-np} with $\ell=1$.  After removing their contributions,
additional singular points $\CA_{\pm (0,\ell)}$ for odd $\ell$ are
revealed.  Their non-perturbative amplitudes are all of the truncated
form \eqref{eq:Fgap-np} and therefore can be removed all together.  In
addition, the contributions from $\CA_{\pm (1,0)}$ as well as
$\CA_{\pm (1,1)}$, whose non-perturbative amplitudes are of the
non-trivial form \eqref{eq:DelF-B}, can also be successfully removed,
given that the correct Stokes constants
\begin{equation}\label{eq:S10S11}
  S_{\pm (1,0)} = S_{\pm (1,1)} = 8
\end{equation}
are chosen.  The Borel plane after successful removal of all these
singular points are given in Fig.~\ref{fig:Nf4tau2i_remove}.

\begin{figure}
  \subfloat[$\imag s_{g}^{(j=-1)}$]{\includegraphics[height=3cm]{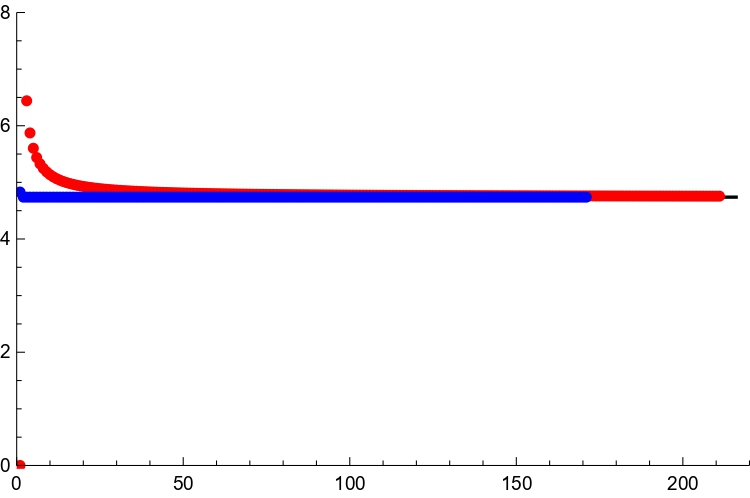}}%
  \hspace{1ex}
  \subfloat[$\imag s_{g}^{(j=0)}$]{\includegraphics[height=3cm]{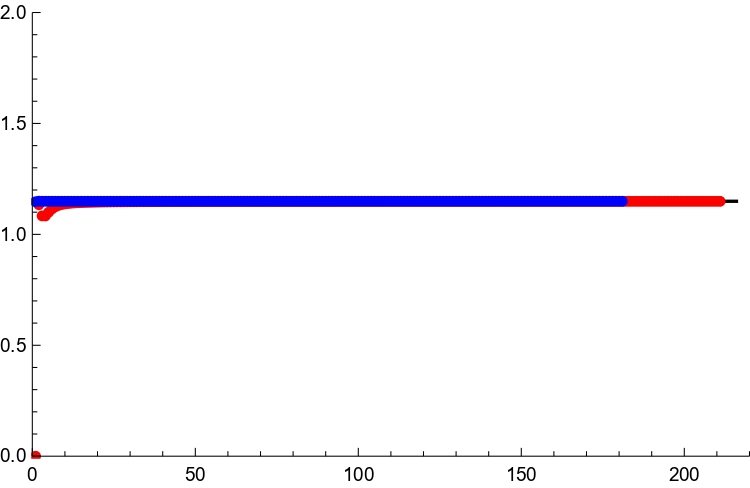}}%
  \hspace{1ex}
  \subfloat[$\imag s_{g}^{(j=1}$]{\includegraphics[height=3cm]{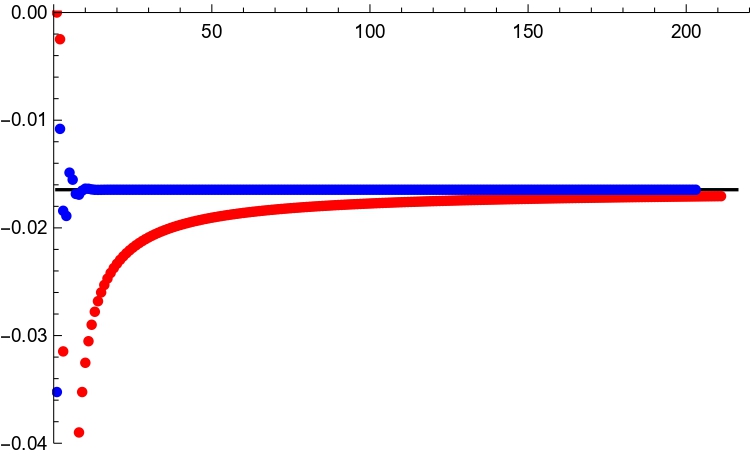}}
  \caption{The sequences $s_g^{(j=-1,0,1)}$ for non-holomorphic free
    energies of $4d$ massless $N_f=4$ SQCD (electric frame) with
    $\tau=3\ri$ and $\tau_2 = 3$.  Red dots are numerical values, blue
    dots are results after Richardson transform, and black lines are the
    expected asymptotic value, the coefficient $a_j^{(\CA_\mu),+}$ of
    non-perturbative amplitude for $\mu=\pm (0,1)$ and $j=-1,0,1$,
    multiplied with Stokes constant $S_{\pm (0,1)}=1$.  Note that
    $s^{(j)}_g$ are all purely imaginary.}
  \label{fig:LOR-Nf4} 
\end{figure}

\begin{figure}
  \centering%
  \subfloat[$\tau=3\ri, \CA_{\pm (0,1)}=\pm 2\pi$ removed]%
  {\includegraphics[height=4.5cm]{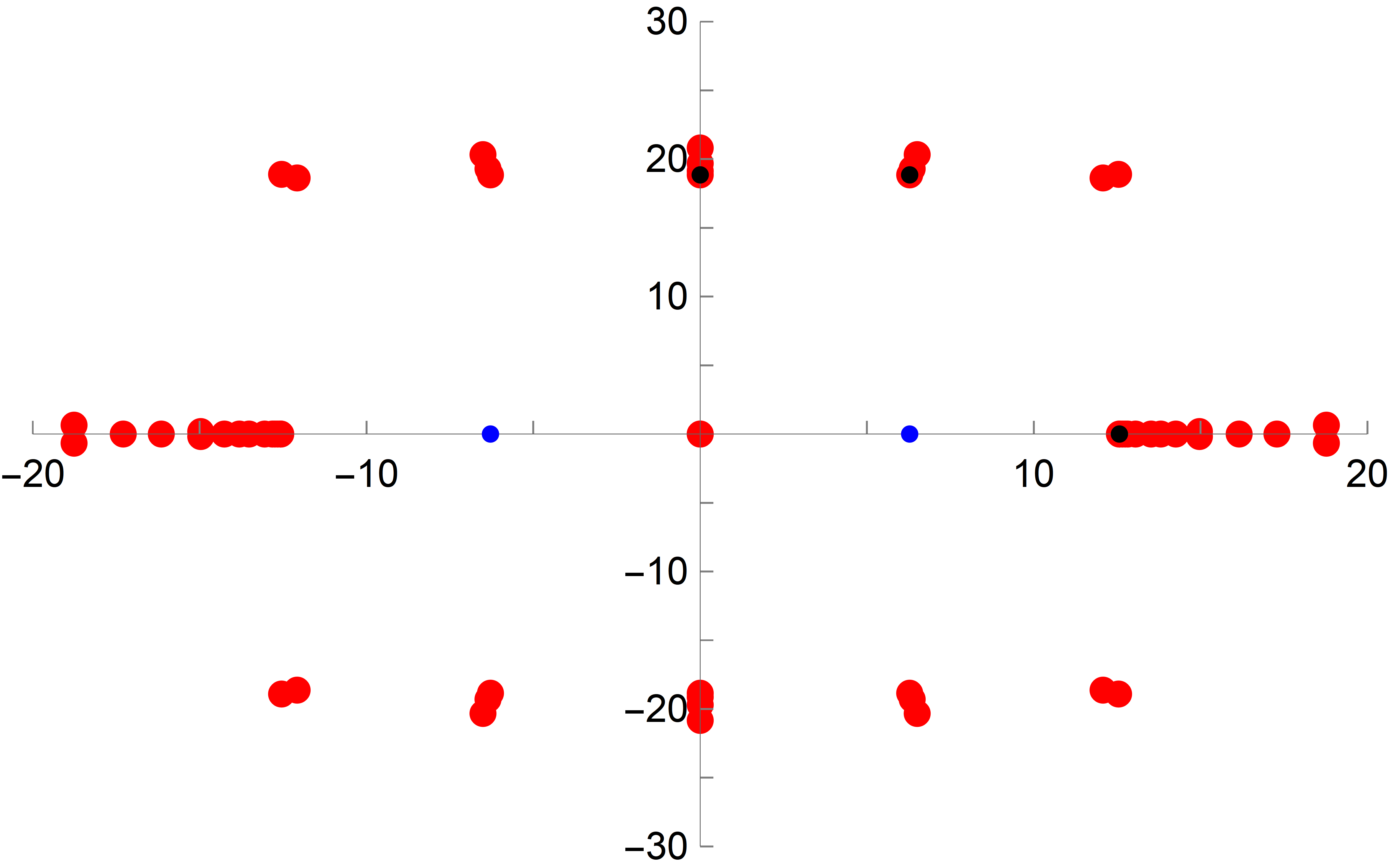}}%
  \hspace{2ex}%
  \subfloat[$\tau=\ri/3, \CA_{\pm (1,0)}=\pm 2\pi\ri/3$ removed]%
  {\includegraphics[height=4.5cm]{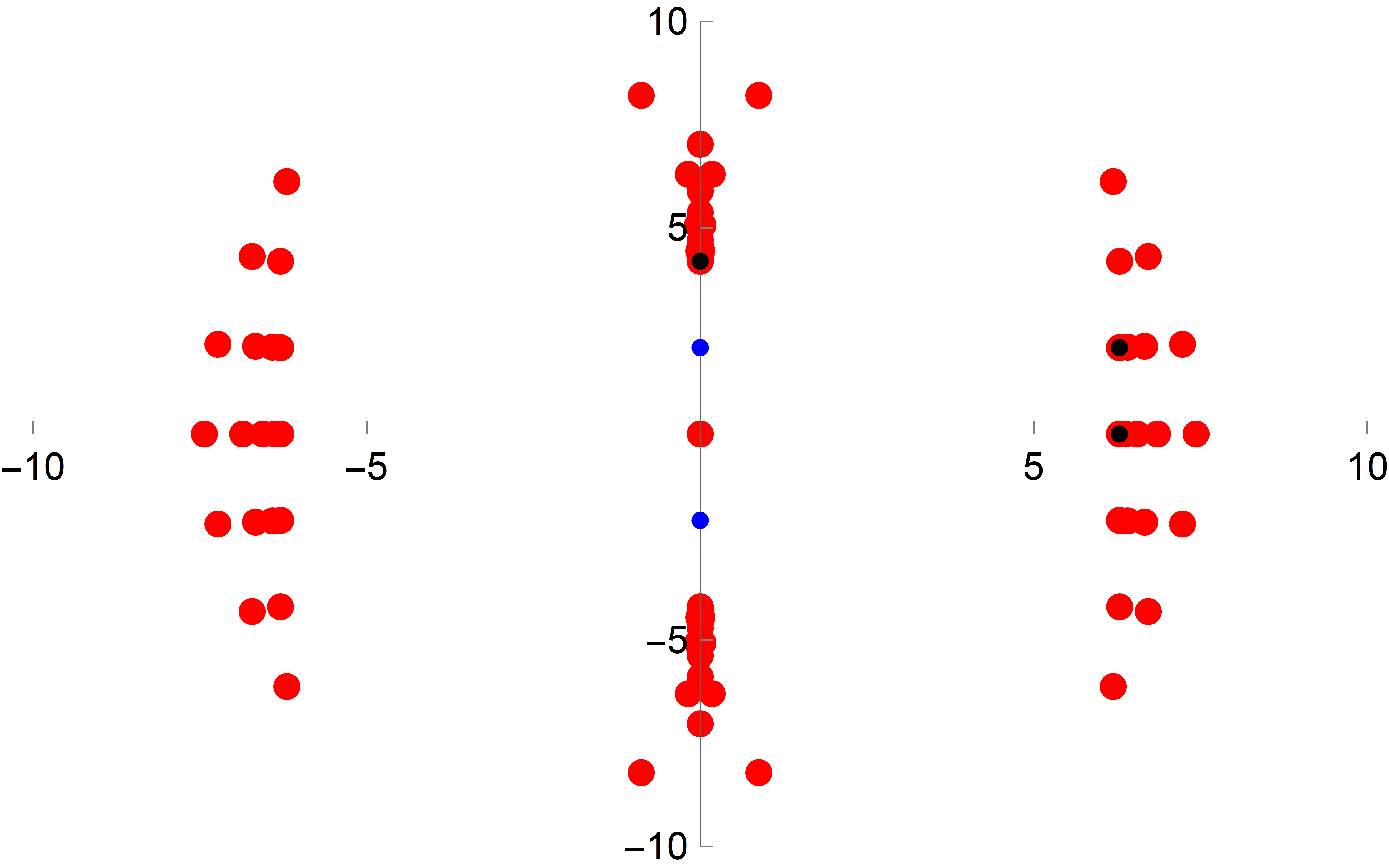}}
  \caption{Borel plane of non-holomorphic perturbative free energies
    (electric frame) of 4d massless $N_f=4$ theory after the dominant
    singularities have been removed, for (a) $\tau = 3\ri$ and
    $\tau_2 = 3$, where the dominant singularities are
    $\CA_{\pm (0,1)}$ (marked in blue), and for (b) $\tau = \ri/3$ and
    $\tau_2 =1/3$, where the dominant singularities are
    $\CA_{\pm (1,0)}$ (marked in blue). The sub-dominant singularities
    are marked in black.}
  \label{fig:brlsub-Nf4}
\end{figure}

Next, we proceed to study the modular form of the non-perturbative
contributions without taking the holomorphic limit.  As shown in
Fig.~\ref{fig:brl-nonholtau3i}, the location of singular points of the
Borel transform is still of the form \eqref{eq:A-gamma-Nf4}.  The
first few non-perturbative coefficients of the dominant singular
points can be checked to agree with \eqref{eq:Fmu},\eqref{eq:SD} by
using the resurgence relation \eqref{eq:rel} and the auxiliary
sequences $s_g^{(j)}$ \eqref{eq:seq-sg}, as shown in
Fig.~\ref{fig:LOR-Nf4}.
Higher order non-perturbative coefficients can be verified by
successful removal of their contributions from the resurgence relation
in $F_g$, so that the corresponding singular points disappear from the
Borel plane, and as shown in Fig.~\ref{fig:brlsub-Nf4}.

Although we have only found Stokes constants corresponding to a subset
of the BPS spectrum, the remainder can be uncovered using the modular
properties of the non-perturbative amplitudes.  Suppose we already
have non-perturbative sectors with charge $\mu=(p,q)$,
non-perturbative amplitude $F^{(\mu)}$, and Stokes constant $S_{\mu}$,
if us move continuously and simultaneously in $\IH$ and $\IH^*$ from
$\tau$ and $\bar{\tau}$ to $\gamma\cdot\tau$ and
$\gamma\cdot\bar{\tau}$ for some $\gamma \in SL(2,\IZ)$, by an
argument similar to those in Section~\ref{sec:mon}, we will uncover an
additional non-perturbative sector with charge $\mu\cdot\gamma$, whose
non-perturbative amplitude is $F^{(\mu\cdot\gamma)}$, and whose Stokes
constant is the same $S_{\mu\cdot\gamma} = S_{\mu}$.  In this way, we
conclude that there are non-perturbative sectors with all the charges
$(p,q)$ and $2(p,q)$ where $(p,q)$ are co-prime integers, and the
associated Stokes constants are
\begin{equation}
  S_{(p,q)} = 8,\;\; S_{(2p,2q)} = -2,
\end{equation}
which agree completely with the BPS spectrum of the 4d massless
$N_f=4$ theory.

\section{Discussion}
\label{sec:dis}

In this paper we study the modular properties of non-perturbative
contributions to topological string free energies.  We find that
non-perturbative contributions are modular covariant with respect to
the monodromy group, a congruent subgroup of the symplectic group or
the modular group of periods of Calabi-Yau threefold, in the sense
that every non-perturbative amplitude $F^{(\mu)}$ is labeled by a
charge vector $\mu$, and a monodromy transformation maps the
non-perturbative amplitude $F^{(\mu)}$ to $F^{(\mu\cdot\gamma)}$.
Furthermore, we find that in each stability chamber of the moduli
space, non-perturbative sectors form orbits of the local monodromy
group generated by loops around singular points of the moduli space
inside the stability chamber, and that the Stokes constants are the
same across the orbits. In the case of topological string models
associated to 4d gauge theories, such as Seiberg-Witten theory, or 4d
massless $N_f=4$ SQCD, infinitely many Stokes constants can be
generated from a small set of data, and they coincide with the full
spectrum of infinitely many BPS invariants of these theories,
providing evidence to the conjecture that Stokes constants of
topological string free energy can be identified with BPS invariants.
These are the first examples other than the simple models such as
local $\IC^3$ or resolved conifold where the full BPS spectrum have
been reproduced from Stokes constants.

We also show that the generators of Stokes transformations of the
non-holomorphic partition function also satisfy Lie brackets of the
Kontsevich-Soibelman Lie algebra, just like the holomorphic case
\cite{Douaud:2026qfo}, supporting the conjecture that the global
Stokes transformation can be identified with the spectrum generator of
BPS invariants, both of which with mild assumptions do not change upon
wall-crossing.

There are various directions to generalize this work.  The most
obvious is to consider compact Calabi-Yau threefolds \cite{Gu:2023mgf}
and study if non-perturbative contributions also enjoy modular
properties with respect to the monodromy group.  This should be
possible, as the same local monodromy invariance of Stokes constants
has been observed in topological string models on Calabi-Yau
threefolds \cite{Gu:2023mgf}.

Furthermore, in addition to closed string free energies, there have
also been attempts to study non-perturbative contributions to open
string free energies in topological string \cite{Grassi:2022zuk}, or
free energies with defects such as Wilson loops \cite{Gu:2024wag}.  It
would be desirable to understand if these non-perturbative
contributions also have nice modular properties, and how they are
organized under monodromy groups.

Finally, the fact that the generators of Stokes transformations for
topological string partition function form a representation of the
Kontsevich-Soibelman Lie algebra\footnote{It is related to the Stokes
  transformation of topological string partition function given in
  \cite{Iwaki:2023rst}.}, with or without taking the holomorphic
limit, is a very interesting observation.  It would be nice to know,
as pointed out in \cite{Douaud:2026qfo}, whether this leads to a
rigorous proof of the conjecture that Stokes constants equal BPS
invariants.

\section*{Acknowledgment}

We thank Alexander Aleksandrov, Daniel Bryan, Simon Douaud, Zhihao
Duan, Alba Grassi, Alexander Hock, Saebyeok Jeong, Rinat Kashaev,
Amir-Kian Kashani-Poor, Marcos Marino, Tomas Reis, Max Schwick for
useful discussions, and in particular Marcos Marino for participation
in the early stage of the project, as well as Simon Douaud and
Amir-Kian Kashani-Poor for collaboration on related projects.  J.G. is
supported by the National Natural Science Foundation of China (General
Program) funding No.~12375062.

\printindex

\bibliographystyle{amsmod}
\bibliography{Nf4}  

\end{document}